\documentclass{aastex63}
\usepackage{xcolor}
\usepackage{soul}
\usepackage{amsmath}
\usepackage{amsfonts}
\usepackage{dsfont}

\newcommand{\disco}{{\sc DISCO}~}

\received{May 2, 2023}
\submitjournal{Astrophys.\,J}

\shorttitle{Circumbinary Disk Dynamics - I}
\shortauthors{Mahesh et al.}
\graphicspath{{./}{figures/}}

\begin{document}

\title{Analytical and Numerical Analysis of Circumbinary Disk Dynamics - I: Coplanar Systems}

\correspondingauthor{Siddharth Mahesh}
\email{sm0193@mix.wvu.edu}

\author{Siddharth Mahesh}
\affiliation{Department of Physics and Astronomy, West Virginia University, Morgantown, WV 26506, USA}
\affiliation{Center for Gravitational Waves and Cosmology, West Virginia University, Chestnut Ridge Research Building, Morgantown, WV 26505, USA}

\author{Sean T. McWilliams}
\affiliation{Department of Physics and Astronomy, West Virginia University, Morgantown, WV 26506, USA}
\affiliation{Center for Gravitational Waves and Cosmology, West Virginia University, Chestnut Ridge Research Building, Morgantown, WV 26505, USA}

\author{Michal Pirog}
\affiliation{Department of Physics and Astronomy, West Virginia University, Morgantown, WV 26506, USA}
\affiliation{Center for Gravitational Waves and Cosmology, West Virginia University, Chestnut Ridge Research Building, Morgantown, WV 26505, USA}

\begin{abstract}

We present an analytical and numerical study of a system composed of a stellar binary pair and a massless, locally isothermal viscous accretion disk that is coplanar to the binary orbital plane. Analytically, we study the effect of the binary's gravitational potential over short timescales through the stability of epicyclic orbits, and over long timescales by revisiting the concept of resonant torques. Numerically, we perform two-dimensional Newtonian simulations of the disk-binary system over a range of binary mass ratios. We find that the results of our simulations are consistent with those of previous numerical studies. We additionally show, by comparison of the analytical and numerical results, that the circumbinary gap is maintained on the orbital timescale through the driving of epicyclic instabilities, and does not depend on resonant torquing, contrary to the standard lore. While our results are applicable to any disk-binary system, we highlight the importance of this result in the search for electromagnetic and gravitational-wave signatures from supermassive black-hole binaries.

\end{abstract}


\section{Introduction}
\label{sec:intro}

The dynamics of gaseous accretion disks under the gravitational influence of binary point masses is a relevant topic in multiple astrophysical contexts, including protoplanetary disks around binary star systems and accretion disks surrounding supermassive black-hole binaries. For an equal-mass system (i.\,e.\,where the masses of the two stars in the binary system are equal), the binary clears a central low-density cavity out to roughly twice the binary's orbital separation \citep{2005ApJ...622L..93M}. In the case of supermassive black-hole binaries, the traditional picture holds that the binary and the cavity co-evolve until the gravitational-wave timescale for the binary drops below the viscous timescale of the inner disk edge, at which point the binary and the disk decouple from each other. This picture is largely motivated by analytical studies \citep{GT1980ApJ, AL94} which suggested that the circumbinary gap is opened and maintained by the balancing of dynamical torques from the binary (which deposit angular momentum at discrete locations known as Lindblad resonances) and viscous torques that dissipate the angular momentum through spiral density waves. However, full numerical simulations of circumbinary accretion disks \citep{DOrazio, late_inspiral} have not been consistent with this analytical picture due to the lack of a demonstrable decoupling phase or agreement in the size of the circumbinary cavity. In addition, in a companion study, we find that the disagreement is more pronounced for disk-binary systems with relative inclinations when we compare an extension of the analytical results presented here with numerical simulations (\cite{paper2}, hereafter Paper II).

Therefore, we revisit the predictions for the circumbinary gap size as presented in \cite{AL94}. We recapitulate the background physics of the system considered in \cite{AL94} and provide a prescription for reproducing their results. Additionally, in order to reconcile the observed inconsistencies, we suggest a new picture of gap opening that is not contingent on resonant torquing and occurs over timescales comparable to the binary period, namely the propagation of dynamical instabilities due to the asymmetric nature of the gravitational potential. We show that this new approach successfully predicts the observed trend in the gap sizes with varying mass ratio as studied in this paper, and lends itself naturally to the case of the inclined disk-binary system which we study in Paper II.

While analytical methods allow for an understanding of the effect of the binary potential on the disk dynamics with minimal inclusion of disk viscosity, numerical studies allow for a full hydrodynamic study of the same system. Previous numerical studies of circumbinary disks were done in \cite{Artymowicz_Lubow_1996} using smooth particle hydrodynamics and in \cite{Gunther} using a grid-based hydrodynamical model with viscosity. In our approach, we build on the setup introduced in \cite{MacFadyen_2008}, where the authors focused on an equal mass binary, and \cite{DOrazio}, where the authors considered a varying binary mass ratio and its effect on the accretion process. To conduct our numerical investigations, we use the \disco code originally described in \cite{disco}, which has been successfully used to solve more advanced problems in circumbinary disk dynamics such as the gap opening criteria \citep{gap_opening_disco}, gas dynamics inside the central cavity \citep{gas_dynamic}, and disk evolution during the late inspiral stage of a supermassive black hole binary \citep{2016MNRAS.459.2379D, late_inspiral}.

Our study is divided into two papers on the basis of the two classes of systems being studied (coplanar and inclined), with each paper providing the requisite analytical and numerical investigations to characterize the system and reconcile the results obtained from both approaches. In this paper, we present the simple, well-studied case of a coplanar binary and circumbinary disk, in order to establish a detailed self-contained reference for the theoretical preliminaries and numerical setup used in this two-part study and future studies. 
In Paper II, we study the case of inclined disk-binary systems, where we note that the results of the numerical simulations further validate the new analytical approach to understanding the clearing of a central gap. 

In this paper, the next section contains the theoretical preliminaries required to motivate the numerical simulations (Section \ref{sec:preliminary_mp}) and analytical computations (Section \ref{sec:preliminary_sid}). Section \ref{sec:nr} briefly describes the numerical code (we leave readers to consult \cite{disco} for further details) and our specific setup, as well as the results from our numerical simulations, focusing on the density and angular velocity distributions of the disk. Section \ref{sec:sid} presents an analytical investigation of disk dynamics by studying the stability of perturbed orbits in the binary potential
and re-examining the resonant torque picture. We introduce two separate definitions for the size of the gap truncated by the binary due to these two descriptions, and present the results of the corresponding gap size computations and a comparison with the numerical results. We summarize and discuss our results in Section \ref{sec:conc}. 

\section{Preliminaries}\label{sec:preliminary}

We assume throughout this work that the self-gravity of the disk is negligible, and there is no back-reaction on the evolution of the binary due to the disk. The binary itself is composed of two point masses, the more massive one referred to as the \textit{primary} and the less massive one as the \textit{secondary}, and these point masses execute fixed circular Keplerian orbits. The disk, on the other hand, is treated either as a collection of massless test particles for the purpose of analytical calculations, or as a viscous, non-self-gravitating, locally isothermal fluid for the purpose of numerical simulations. This description of the binary is valid in the regime where the binary separation is large
relative to the gravitational radii of the two masses, so that the binary does not evolve over the course of the simulation.
Even if the binary executes a slow, adiabatic inspiral through the emission of gravitational waves, the orbital separation only changes over a timescale much larger than the orbital period or the viscous timescale, so for our purpose it can be treated as a constant. These two physical treatments approach the dynamics of the disk over different timescales; the analytical treatment addresses the dynamics over timescales comparable to the binary orbital period, whereas the numerical treatment addresses the large scale structure and long term evolution of the disk, which require a full hydrodynamic treatment. The large orbital separations also justify the use of Newtonian physics for the evolution of the overall system.

The binary components are point masses moving in circular, Keplerian orbits in the center-of-mass frame, with the disk moving in the combined Newtonian gravitational potential of both masses. We will see analytically that, for gas orbiting far enough from the binary, the combined binary potential can be expressed as small, periodic perturbations about the potential of a single point mass whose mass equals the total binary mass, and which is at the location of the binary center of mass. The disk is assumed to be geometrically thin and not self gravitating, justifying a Keplerian rotation law about the center of mass with no additional contribution to the potential due to the density distribution of the gas. Numerically, this rotation law will serve, modulo small corrections, as a leading term in the disk's initial angular velocity distribution. The disk is viscous, with the viscosity following an $\alpha$-prescription. We work in geometrized units throughout, where the universal gravitational constant, $G$, and the speed of light, $c$, satisfy $G=c=1$; we also set the total mass of the binary system, $M$, and the orbital separation of the binary, $a$, to unity. In the following subsections, we provide further details of the disk and binary dynamics.  

\subsection{Hydrodynamics in the Binary Potential} \label{sec:preliminary_mp}

We work in standard two-dimensional polar coordinates $(r, \phi)$, and we use the open-source numerical code \disco \citep{disco} to numerically solve the Riemann problem for the four primitive variables: surface mass density $\sigma(\mathbf{x},t)$, surface pressure $p(\mathbf{x},t)$, and radial and azimuthal components of the velocity $\mathbf{v}(\mathbf{x},t) = (v^r(\mathbf{x},t), v^\phi(\mathbf{x},t))$. We suppress the functional dependence on $(\mathbf{x},t)$ hereafter, and refer to \cite{disco} for the detailed description of the numerical methods. The Newtonian hydrodynamic field equations can be formulated in the standard ``primitive'' form typically found in textbooks, or in the ``conservative'' form typically employed when finding numerical solutions. We employ the latter, and will briefly review the basic equations here. The continuity equation and Euler equation read \citep{accrition_power}
\begin{eqnarray}\label{hydro}
\partial_t \sigma + \nabla \cdot (\sigma \mathbf{v}) & = & 0, \\
\partial_t \mathbf{v} + \mathbf{v} \cdot \nabla \mathbf{v} & = & \frac{1}{\sigma} \nabla p - \nabla \Phi + \mathbf{f}_\nu
\end{eqnarray}
$\Phi$ is the total gravitational potential due to both test masses,
\begin{equation}
\Phi(\mathbf{x}) = \frac{Gm_1}{|\mathbf{x} - \mathbf{x_1}|}+\frac{Gm_2}{|\mathbf{x} - \mathbf{x_2}|},
\end{equation}
where $m_1$ and $m_2$ are the masses of the binary components, and $\mathbf{x_1}$ and $\mathbf{x_2}$ are their positions on fixed circular orbits.
$\mathbf{f}_\nu$ describes the force due to viscosity $\nu$,
\begin{equation}
\mathbf{f}_\nu = \nabla \cdot \nu \nabla \mathbf{v} + \nabla \left( \frac{1}{2} \nu \nabla \cdot \mathbf{v} \right).
\end{equation}
The system of coordinates $(r,\phi)$ lies in the plan of the disk and is centered on the binary’s center of mass. We complete this system of equations using the $\alpha$ prescription for viscosity \citep{sha-sun},

\begin{equation}\label{alpha_type}
\nu(r) = \alpha s^2 r^{3/2}
\end{equation}
and a locally isothermal equation of state,
\begin{eqnarray}\label{eos}
p &=& s^2 \sigma\,, \\ \nonumber
s(r) &=& \chi / \sqrt{r}\,,
\end{eqnarray}
where $s$ is the speed of sound and $\chi = h/r$ is the constant scale height-to-radius ratio. While this study primarily sets the free parameter of $\alpha = 0.01$, we ran simulations with $\alpha \in \{0.001, 0.003, 0.01, 0.03, 0.1\}$, in order to assess the generality of our outcomes. We set the free parameter $\chi = 0.1$ in all of our simulations.
\subsection{Test Particles in the Binary Potential} \label{sec:preliminary_sid}

In our units where $G=M=a=1$, the only remaining free parameter is the mass ratio of the point masses, which
we choose to express as the secondary mass ratio $\mu \equiv m_2/M$,
where $m_2$ is the mass of the secondary.
In the center-of-mass frame, the locus of the primary, $\mathbf{x_1}$, and the secondary, $\mathbf{x_2}$, are given in two-dimensional Cartesian coordinates by
\begin{eqnarray}\label{eqn:binarymotion}
    \mathbf{x_1}(t) & = & \mu(\cos t,\sin t), \\ \nonumber
    \mathbf{x_2}(t) & = & (\mu-1)(\cos t,\sin t).
\end{eqnarray}
We note that the binary orbital frequency $\Omega_b = \sqrt{\frac{GM}{a^3}}$ has a magnitude of unity due to our choice of normalization,
so that the orbital frequency of the binary is simply $t$ in these units.
The corresponding Newtonian gravitational potential, $\Phi$, experienced by a test particle located at $\mathbf{x}$ is given by
\begin{equation}\label{eqn:truepotential}
    \Phi = -\frac{1-\mu}{|\mathbf{x} - \mathbf{x_1}|} - \frac{\mu}{|\mathbf{x} - \mathbf{x_2}|}.
\end{equation}
With some algebra, we can write the potential in two-dimensional polar coordinates $(r,\phi)$. 
In this form, one can exploit the generating function of Legendre polynomials as in \cite{2014grav.book.....P} to obtain the multi-polar decomposition
\begin{equation}\label{eqn:multipoles}
    \Phi = -\sum^\infty_{l=0}\mathcal{Q}_l r^{(-l-1)} P_l(\cos(\phi - t)),
\end{equation}
where $\mathcal{Q}_l \equiv (-\mu)^l(1-\mu) + \mu(1 - \mu)^l$ is the dimensionless multipole moment and $P_l$ is the Legendre polynomial of order $l$. We separate the monopole ($l=0$) term, which we will refer to as $\Phi_\mathrm{M}$ and is given by
\begin{equation}\label{potential_M}
    \Phi_\mathrm{M}(r) = -\frac{1}{r}.
\end{equation}
$\Phi_\mathrm{M}$ is static and spherically symmetric, and can be considered the background potential.
We then bundle all other multipole terms into a single perturbing potential $\Phi_\mathrm{P}$, to indicate that the effect of all higher multipoles will enter the subsequent perturbation theory treatment at the same order. We also note that all higher multipoles manifest as harmonics of the variable $(\phi - t)$, and it is therefore convenient to re-express $\Phi_\mathrm{P}$ as a Fourier series of the form
\begin{equation}
    \Phi_\mathrm{P} = - \sum_{l=2}^{\infty}\mathcal{Q}_l r^{(-l-1)}\mathcal{P}_l(\cos(\phi-t)) = \sum_{m = 2}^{\infty} \Phi_m \cos\left(m(\phi - t)\right) .
\end{equation}
We begin our summation from $l = 2$, since the dipole moment $\mathcal{Q}_1=0$. The expression for the Fourier coefficients, $\Phi_m$, can be found in Equation (16) of \cite{2015MNRAS.452.2396M} for the general case of an inclined, eccentric binary source. For the purposes of this paper, we set the inclination and eccentricity to zero. Subsequent papers will tackle the modifications to the physics introduced by these additional parameters both in the analytics and the numerics.

The Hamiltonian, $\bar{\mathcal{H}}$, for a test particle evolving under these potentials takes the form
\begin{equation}
    \bar{\mathcal{H}} = \frac{1}{2}\left(p_r^2 + \frac{p_\phi^2}{r^2}\right) + \Phi_\mathrm{M} + \Phi_\mathrm{P},
\end{equation}
where $p_r$ and $p_\phi$ are the canonical momenta corresponding to the position variables $R$ and $\varphi$, respectively. We then perform a canonical transformation to a frame of reference still centered on the center-of-mass, but now co-rotating with the binary. The transformed coordinates
are related to the original coordinates as
\begin{equation}
    r \rightarrow r \quad ; \quad \phi \rightarrow \varphi = \phi -   t.
\end{equation}
To obtain the corresponding transformation for the momenta, we invoke a generating function of the second kind 
\begin{eqnarray}
    F_2 & = & rp_r + (\phi -   t)p_\phi, \\ \nonumber
    (r , \varphi) & = & \left(\frac{\partial F_2}{\partial p_r} , \frac{\partial F_2}{\partial p_\phi}\right).
\end{eqnarray}
We can see that the transformed momenta are identical since
\begin{eqnarray}
    p_r & = & \frac{\partial F_2}{\partial r} = p_r ,\\
    p_\varphi & = & \frac{\partial F_2}{\partial \phi} = p_\phi.
\end{eqnarray}
Finally, the new Hamiltonian is given by
\begin{equation}
    \mathcal{H} = \bar{\mathcal{H}} + \frac{\partial F_2}{\partial t}.
\end{equation}
Thus, the final Hamiltonian, after plugging in the value of the potentials in this frame, is given by
\begin{equation}
    \mathcal{H} = \frac{1}{2}\left(p_r^2 + \frac{p_\varphi^2}{r^2}\right) -   p_\varphi - \frac{1}{r} - \sum_{m=2}^{\infty}\Phi_m\cos(m\varphi).
\end{equation}

Before we solve the corresponding Hamilton's equations perturbatively, we state the full equations of motion of this Hamiltonian system as
\begin{eqnarray}\label{fullEOM}
    \dot{r} & = & p_r,\\
    \dot{\varphi} & = & \frac{p_\varphi}{r^2} - 1,\\
    \dot{p}_r & = & \frac{p_\varphi^2}{r^3} - \frac{1}{r^2} - \Phi'_m\cos(m\phi),\\
    \dot{p}_\varphi & = & m\Phi_m\sin(m\varphi)
\end{eqnarray}
 For brevity, we use the notation $\dot{x}$ to denote the time derivative of $x$, $\Phi_m'$ to denote the derivative of the potential mode $\Phi_m$ with respect to the radial position $r$. Unless stated otherwise, there is an implied summation over the Fourier mode label $m$ wherever $\Phi_m$ appears We assume a Keplerian rotation profile instead of including the axisymmetric components of the higher multipoles, because they represent negligible $\mathcal{O}(r^{-l})$ corrections to the Keplerian rotation profile and, as a result, to the locations of the resonances and other results. It should be noted that, while the corrections become non-negligible at radii closer to the binary separation, the perturbation theory treatment breaks down in those regions and the dynamics is no longer linear. The perturbation theory treatment follows from the derivation of Lindblad resonances in \cite{BinneyTremaine}.
 
We consider perturbations about a circular orbit at $r = r_0$, the background solution for which is given by
\begin{eqnarray}
    r & = & r_0,\\ \nonumber
    p_r & = & 0,\\ \nonumber
    \varphi & = & \varphi_0 = \omega t,\\ \nonumber
    p_\varphi & = & l_0 = \sqrt{r_0} ,\\ \nonumber
    \omega & = & \sqrt{\frac{1}{r_0^3}} - 1.
\end{eqnarray}
The variable $\omega$ represents the angular frequency of the circular orbit at $r_0$ in the co-rotating frame. The reason for choosing circular orbits is due to the fact that previous numerical studies to find stable test particle orbits under the binary potential have returned either circular or near-circular orbits with a maximum eccentricity, $e$, of $\mathcal{O}(10^{-1})$ (\textit{cf}. Table 1 in \cite{RP1981} and Table 4 of \cite{2005MNRAS.359..521P}). Such small eccentricities can be attributed to the apparent eccentricity introduced by the perturbative effects of the potential $\Phi_P$. 

We now expand about this circular orbit by introducing new variables, given in terms of the canonical coordinates as $(r,p_r,\varphi,p_\varphi) \rightarrow (r_0,0,\omega t,l_0) +(r_1,p_1,\varphi_1,l_1)$. The subscript ``$1$'' indicates that these variables are first order perturbations to the background results, the same order at which the perturbing potential enters the equations of motion. After plugging the new variables into Eq.\,\eqref{fullEOM} and canceling the background terms, we combine the radial position and momentum equations into one second order differential equation. As we will see in Sec.\,\ref{sec:sid}, the only place where $\varphi$ is needed is in building the stability matrix. However, $\varphi$ only enters inside the cosines of $\Phi_P$, which are already first order perturbations. Thus, we can simplify $\varphi = \varphi_0$ and ignore the evolution of $\varphi_1$.  The equations of motion then reduce to
\begin{eqnarray}\label{pertEOM}
    \dot{l}_1 & = & m\Phi_m\sin(m\omega t),\\
    \ddot{r}_1 + (1+\omega)^2r_1 & = & \left(-\Phi'_m -\frac{2(1 + \omega)}{\omega}\frac{\Phi_m}{r_0}\right)\cos(m\omega t). 
\end{eqnarray}
We solve for the angular momentum, $l_1$, by integrating the first equation to give
\begin{equation}\label{epicycle::L}
    l_1 = -\frac{\Phi_m}{\omega}\cos(m\omega t).
\end{equation}
The equation of motion for $r_1$ resembles a driven harmonic oscillator and has the standard solution
\begin{equation}\label{epicycle::R}
    r_1 = -\frac{\cos(m\omega t)}{\left| (1 + \omega)^2 - m^2\omega^2 \right|}\left(\Phi'_m +\frac{2(1 + \omega)}{\omega}\frac{\Phi_m}{r_0}\right)\,.
\end{equation}

We have thus obtained an analytical solution for the radial and angular momentum evolution in the epicyclic approximation. We note immediately that we have \textit{Lindblad resonances} whenever the denominator $(1 + \omega)^2 - m^2\omega^2$ vanishes and the epicyclic amplitude diverges as a result. While these resonances are located at discrete radii, the breakdown of the epicyclic approximation happens over a wider region surrounding these resonances. We shall study the nature of this breakdown and its implications to orbital stability in further sections. 

\section{Numerical calculations} \label{sec:nr}
\subsection{Numerical setup and data analysis}\label{sec:setup}
The numerical code \disco \citep{disco} is used here as a moving mesh, two-dimensional \textit{Harten-Lax-van Leer-Contact (HLLC)} solver \citep{toro} for Newtonian hydrodynamic field equations. Our numerical domain is a flat ring-shaped surface which is centered on the center of mass of the binary and spans in the radial direction from $a$ to $100a$. The central circle-shaped region of radius $a$ where the binary is located is excluded from the computational domain. The numerical grid is composed of 480 concentric radial rings of varying widths. The width of the $n$-th ring is calculated as the difference between two neighboring nodes $\Delta r_n = r_{n+1}-r_n$, where
\begin{equation}\label{node}
r_n = 1 + A \sinh{(BC ) }
\end{equation}
where
\begin{equation}\label{nodeex}
A = \frac{5}{\sinh{(1)}}, \ \ B = \frac{n-1}{480}, \ \ C = \sinh^{-1}{(99/A)} 
\end{equation}

The size of the innermost ring is $\Delta r_1 \approx 0.034$, and the outermost ring is $\Delta r_{480} \approx 0.790$. Each radial ring is composed of a different number of azimuthal zones, which are refined or de-refined as needed to keep the aspect ratio $r \Delta \phi/\Delta r$ close to unity. According to \cite{DRS2011} and \cite{DM2012}, this choice allows the density waves excited by the binary potential to be well resolved over their wavelength $\sim 0.2\pi r$. The time step $\Delta t$ is set to be half of the shortest propagation time across any cell in the grid, i.\,e.\, the width of the cell divided by the local sound speed. The mesh rotates according to $\Omega_\mathrm{grid} = r^{-3/2}$ to minimize diffusive advection errors.

For simplicity, we express all timescales in terms of the binary revolution period $P$. We add that, in our units, $P = 2\pi$. To present the data as a simple function of the distance from the rotation axis, we use two different kinds of averaging. For any quantity $\xi(r, \phi, t)$ we introduce the following notation:
\begin{equation}
\bar{\xi} (r,t) = \frac{1}{2\pi} \int^{2\pi}_0 \xi(r,\phi,t) \ d\phi,
\end{equation}
\begin{equation}
\hat{\xi}(r,\phi) = \frac{1}{\Delta} \int^{t+\Delta }_t \xi(r,\phi,t) \ dt,
\end{equation}
and finally
\begin{equation}\label{average}
\langle \xi \rangle (r)  = \frac{1}{2\pi \Delta} \int^{t+\Delta }_t \int^{2\pi}_0 \xi(r,\phi,t) \ d\phi \ dt. 
\end{equation}
In our simulations we use $\Delta \equiv 50 P$. We dump grid data every $6^\circ$ of the binary revolution, so the time averages
over the interval $\Delta$ are calculated from 3000 instantaneous snapshots of the grid values.

\subsection{Initial data and boundary conditions}\label{sec:id}

We start the simulations with the initial density
\begin{equation}
\sigma(r,t=0) = \sigma_0 \left(\frac{r_\mathrm{s}}{r}\right)^{3} \exp\left[-\left(\frac{r_\mathrm{s}}{r}\right)^{2}\right],
\end{equation}
where $\sigma_0$ is an arbitrary normalization constant, and $r_\mathrm{s}$ is a constant which sets the radial scale of $\sigma$.
The function arguments $(r,t=0)$ are implied throughout the rest of this section. We set $r_\mathrm{s} \equiv 10$, so that the maximum initial value of $\sigma$ occurs at $r = 10\sqrt{2/3} \approx 8.16$.  The initial pressure is calculated from the density using the equation of state, Eq.\,\eqref{eos}.
The initial velocity distribution is given by \cite{MacFadyen_2008} as
\begin{equation}
\mathbf{v} = (v^r, v^\phi) = \left\{ \frac{2}{r \sigma v^\phi } \frac{\partial}{\partial r} \left( r^3 \sigma \nu \frac{\partial}{\partial r} \frac{ v^\phi}{r} \right) , \ \ \left[ \Omega_\mathrm{K}^2 \left( 1+\frac{3}{4} \frac{a^2}{r^2} \frac{q}{(1+q)^2} \right)^2 + \frac{1}{r\sigma} \frac{dp}{dr} \right]^{1/2} \right\}
\end{equation}
where $q \equiv m_1/m_2$ is the \textit{mass ratio}. We calculate $v^\phi$ first and use it to calculate $v^r$, as described in \cite{MacFadyen_2008}.

Boundary conditions are imposed on the primitive variables at both the inner and the outer edges of the grid. The outer boundary conditions for all primitive variables are fixed at the initial data values for the two outermost rings. The same procedure is applied to the components of the fluid velocity at the inner boundary. The exceptions are the density and pressure boundaries, which instead are given by
\begin{eqnarray}
\sigma_i & = & \bar{\sigma}_{i+1} (r_{i+1} / r_i)^{-1/2}, \\
p_i & = & \bar{p}_{i+1} (r_{i+1} / r_i)^{-1/4},
\end{eqnarray}
where $i\in\{1,2\}$ corresponds to the number of each of the two innermost radial rings. We note that these boundary conditions do not prevent the flow of matter into or out of the numerical domain. However, the central excised region does not contain any significant density throughout the entirety of the simulations. We have investigated the accretion rate thought the grid boundaries and the total mass of the system as a function of time, and confirmed that the flow of matter across the boundaries does not impact our results.

\subsection{Mass density and torque distributions} \label{sec:density}

\begin{figure}
\includegraphics[width=0.495\linewidth]{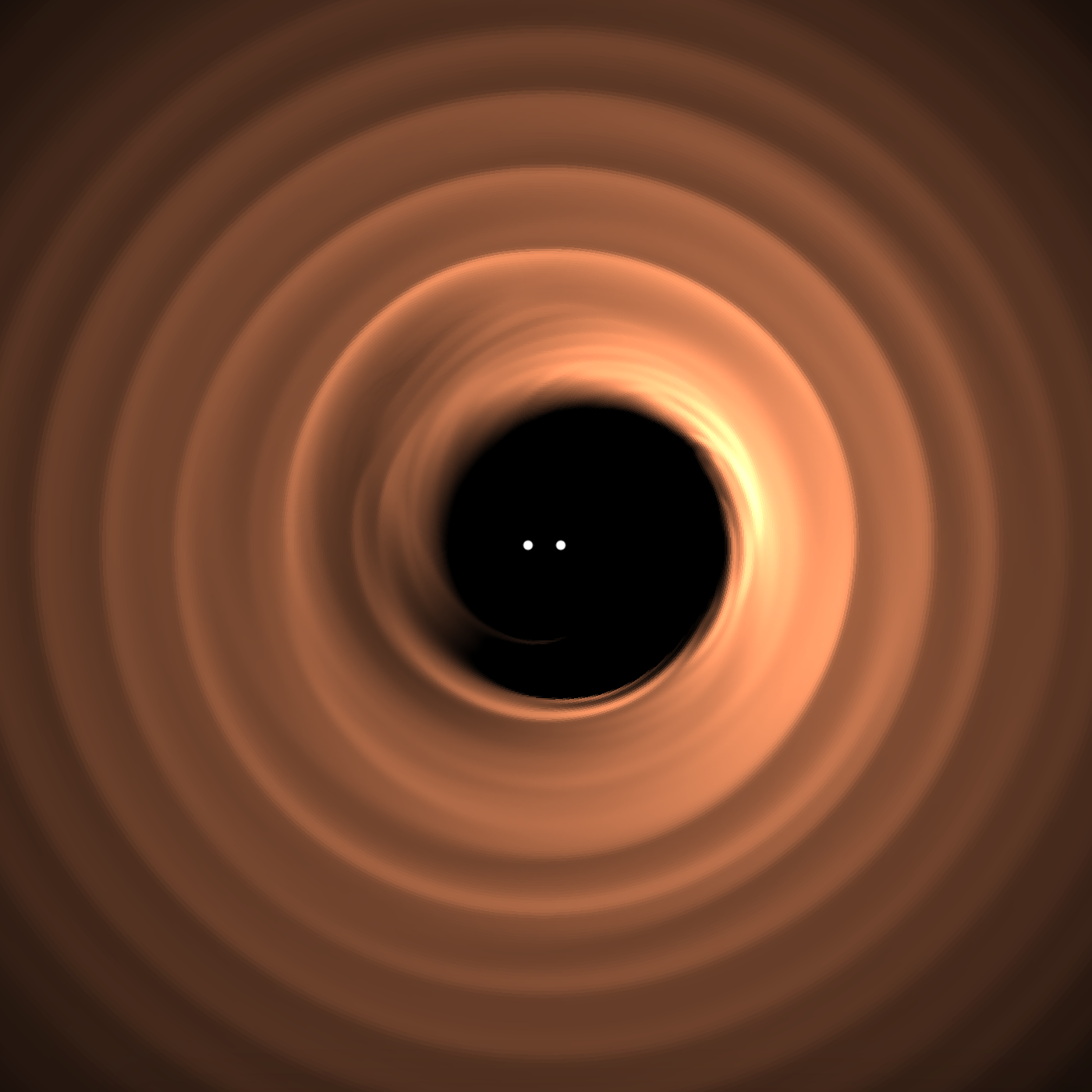}
\includegraphics[width=0.495\linewidth]{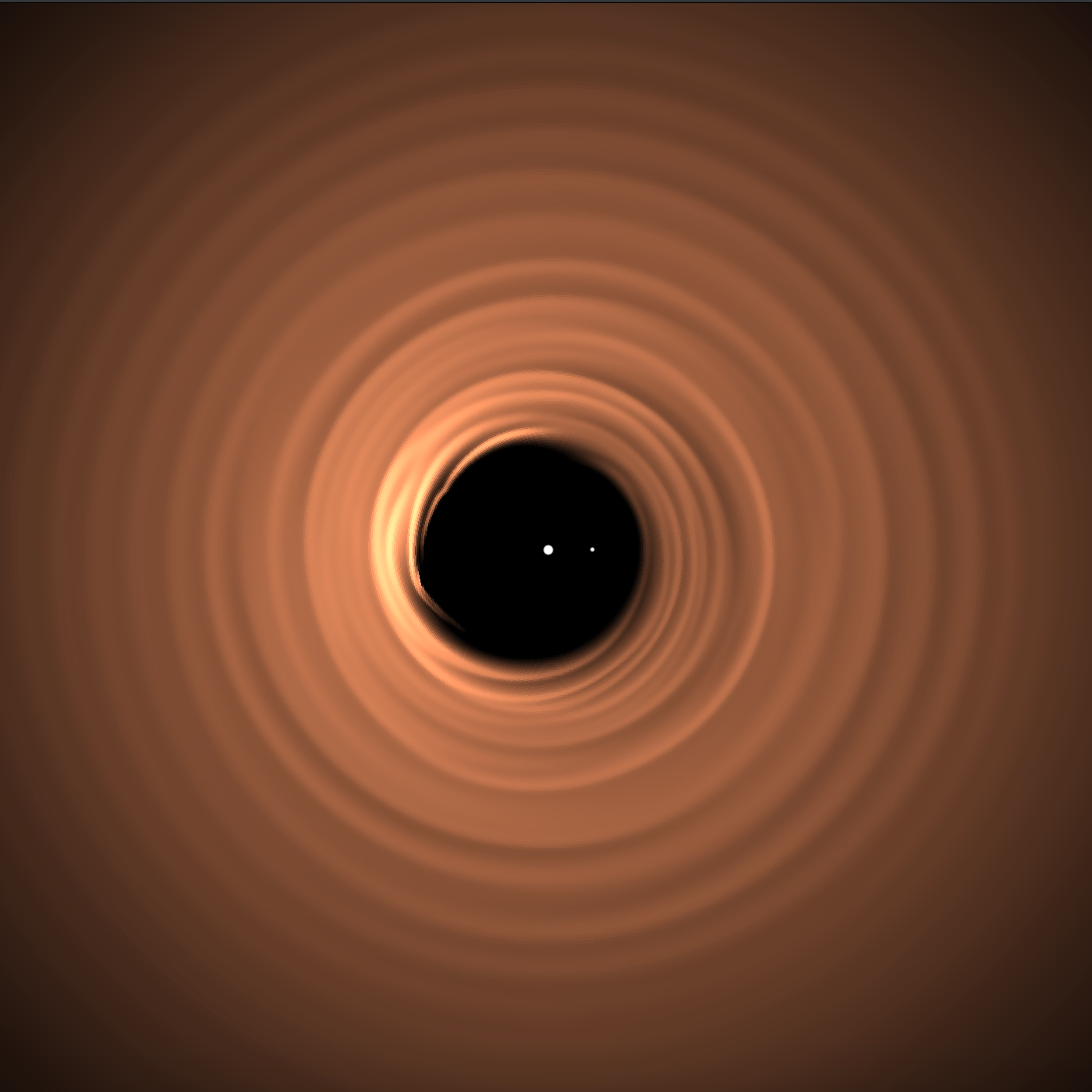}
\includegraphics[width=1.0\linewidth]{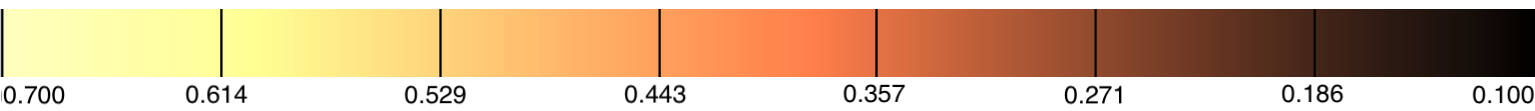}
\caption{Surface mass density $\sigma/\sigma_0$ at $t = 4000 P$ for $q=1$ (left) and $q=1/10$ case (right). The pictures show the region $-15r/a \leq (x,y) \leq 15r/a$.}
\label{fig:2d}
\end{figure}

We have simulated the different binary mass ratios $q$ and viscosity parameters $\alpha$ given in Table \ref{tab:1} until all of the systems settled into a quasi-steady state. That is, until the averaged values (in the sense of Eq.\,\eqref{average}) of the hydrodynamic variables are no longer evolving secularly. Unless stated otherwise, all figures present data for $\alpha = 0.01$. Figure \ref{fig:2d} presents the mass density in the central region of two of these systems. In the left panel,
we present the equal mass ($q=1$) case, and on the right, the $q=1/10$ case.

\begin{deluxetable}{c c c c c c c c c }
\tablecaption{Summary of all simulated configurations. \label{tab:1}}
\tablehead{ $q$ & $\mu=m_2/M$ & $\alpha$ & $t_*$ & $r_\mathrm{dT}$ & $r_\mathrm{max}$ & $r_\mathrm{10\%}$}
\startdata
1:1     &  0.5      & 0.1   & 400   & 1.818  & 3.333 & 1.655   \\
2:3     &  0.4      & 0.1   & 400   & 1.820  & 3.295 & 1.641   \\
3:7     &  0.3      & 0.1   & 400   & 1.800  & 3.103 & 1.606   \\
1:4     &  0.2      & 0.1   & 400   & 1.713  & 2.952 & 1.473   \\
1:10    &  0.(09)   & 0.1   & 400   & 1.561  & 2.766 & 1.322   \\
1:100   &  0.(0099) & 0.1   & 400   & 1.139  & 4.059 & 1.133   \\ \hline
1:1     &  0.5      & 0.03  & 1250  & 1.975  & 3.893 & 2.012   \\
2:3     &  0.4      & 0.03  & 1250  & 1.970  & 3.893 & 1.999   \\
3:7     &  0.3      & 0.03  & 1250  & 1.948  & 3.934 & 1.946   \\
1:4     &  0.2      & 0.03  & 1250  & 1.898  & 3.609 & 1.830   \\
1:10    &  0.(09)   & 0.03  & 1250  & 1.725  & 3.256 & 1.571   \\
1:100   &  0.(0099) & 0.03  & 1250  & 1.251  & 4.101 & 1.232   \\ \hline
1:1     &  0.5      & 0.01  & 4000  & 2.079  & 4.757 & 2.307 \\
2:3     &  0.4      & 0.01  & 4000  & 2.074  & 4.711 & 2.296 \\
3:7     &  0.3      & 0.01  & 4000  & 2.052  & 4.532 & 2.242 \\
1:4     &  0.2      & 0.01  & 4000  & 1.993  & 4.228 & 2.146 \\
1:10    &  0.(09)   & 0.01  & 4000  & 1.862  & 3.411 & 1.828 \\
1:100   &  0.(0099) & 0.01  & 4000  & 1.402  & 3.103 & 1.285 \\ \hline
0       &  0        & 0.01  & ?     & -      & 4.357 & 1.207 \\ \hline
1:1     &  0.5      & 0.003 & 9000  & 2.148  & 5.033 & 2.603   \\
2:3     &  0.4      & 0.003 & 9000  & 2.151  & 4.987 & 2.557   \\
3:7     &  0.3      & 0.003 & 9000  & 2.127  & 4.757 & 2.470   \\
1:4     &  0.2      & 0.003 & 9000  & 2.074  & 4.532 & 2.388   \\
1:10    &  0.(09)   & 0.003 & 9000  & 1.957  & 3.811 & 2.075   \\
1:100   &  0.(0099) & 0.003 & 9000  & 1.518  & 2.620 & 1.388   \\ \hline
1:1     &  0.5      & 0.001 & 24000 & 2.190  & 5.176 & 2.800   \\
2:3     &  0.4      & 0.001 & 24000 & 2.201  & 5.128 & 2.798   \\
3:7     &  0.3      & 0.001 & 24000 & 2.206  & 4.940 & 2.726   \\
1:4     &  0.2      & 0.001 & 24000 & 2.129  & 4.757 & 2.636   \\
1:10    &  0.(09)   & 0.001 & 24000 & 2.040  & 4.143 & 2.297   \\
1:100   &  0.(0099) & 0.001 & 24000 & 1.600  & 2.915 & 1.508   \\ \hline
\enddata
\tablecomments{Columns label the mass ratio $q$, secondary mass ratio $\mu$, viscosity $\alpha$, time at which we report results (i~e.~after the system reaches a quasi-steady state) $t_*$, radius where the torque densities balance $r_\mathrm{dT}$, radius of the maximum density $r_\mathrm{max}$, and radius where the density inside the cavity falls to $10\%$ of the final density maximum $r_\mathrm{10\%}$.}
\end{deluxetable}

\begin{figure}
\begin{center}
\includegraphics[width=0.6\linewidth]{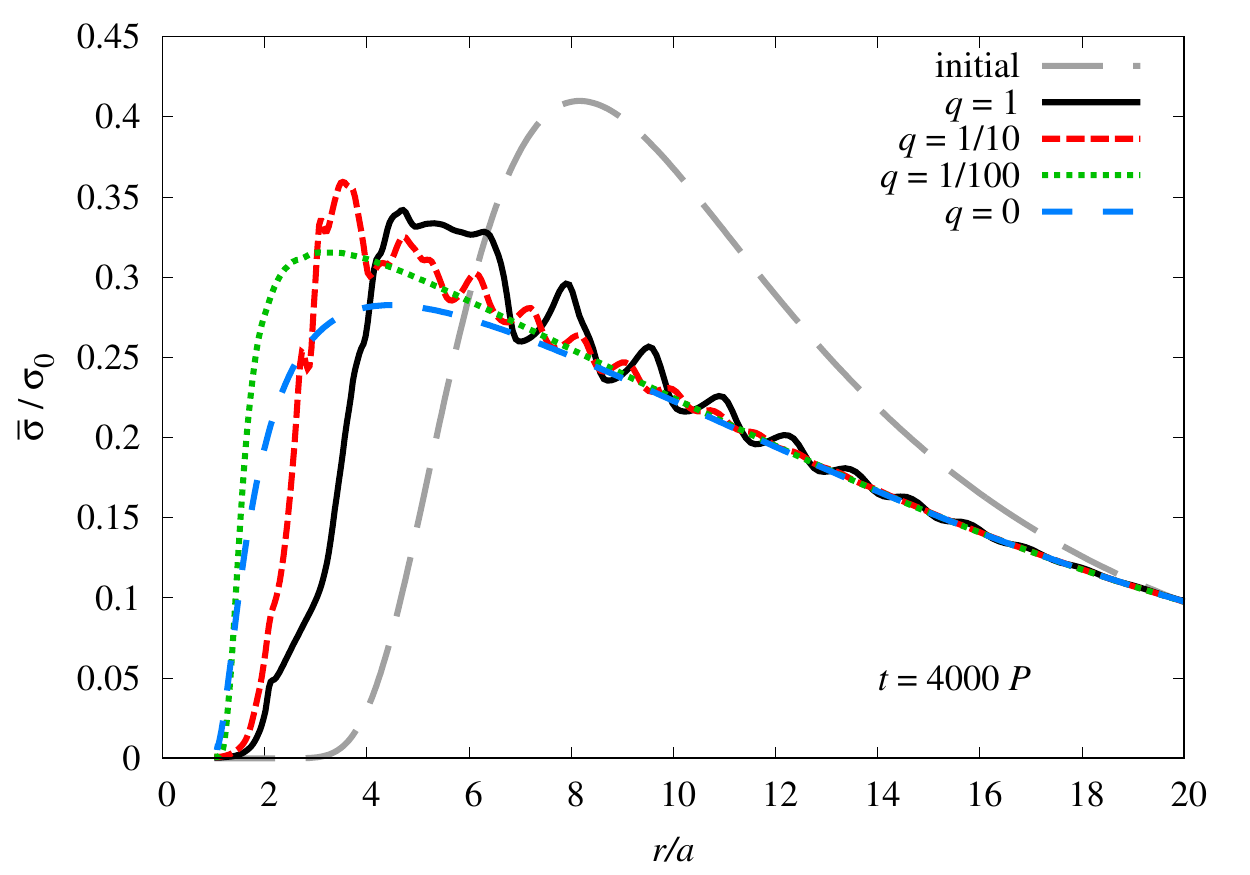}
\caption{Azimuthally averaged surface mass density, at $t=4000
 P$ for $q \in \{0, 1/100, 1/10, 1 \}$ together with the initial density for comparison.}
\label{fig:not_avg}
\end{center}
\end{figure}

\begin{figure}
\includegraphics[width=\linewidth]{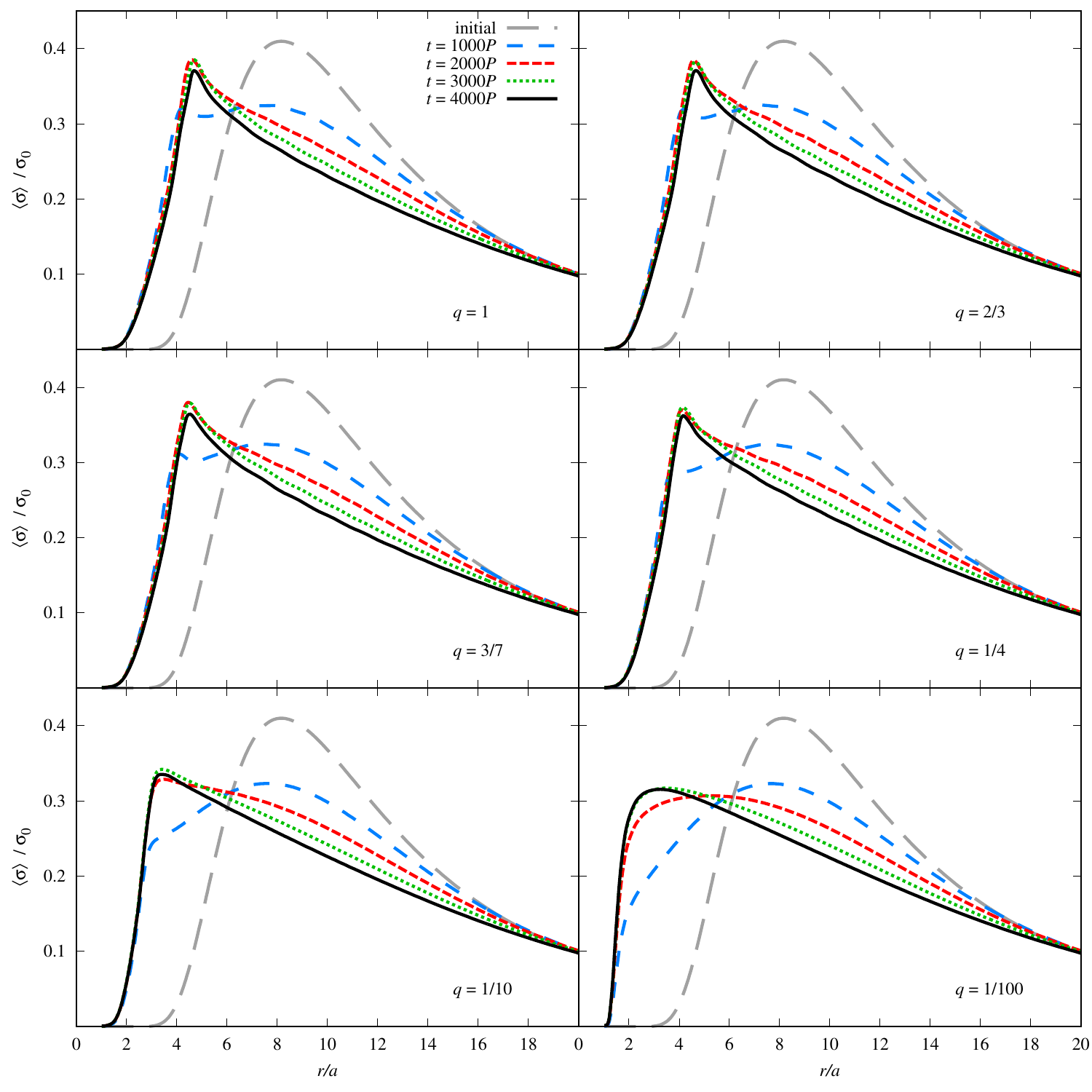}
\caption{Averaged surface mass density initially and at four different times $t \in \{1000 P, 2000 P, 3000 P, 4000 P\}$, with each panel considering one of the mass ratios summarized in Table \ref{tab:1}.} 
\label{fig:dens_sep}
\end{figure}

\begin{figure}
\includegraphics[width=\linewidth]{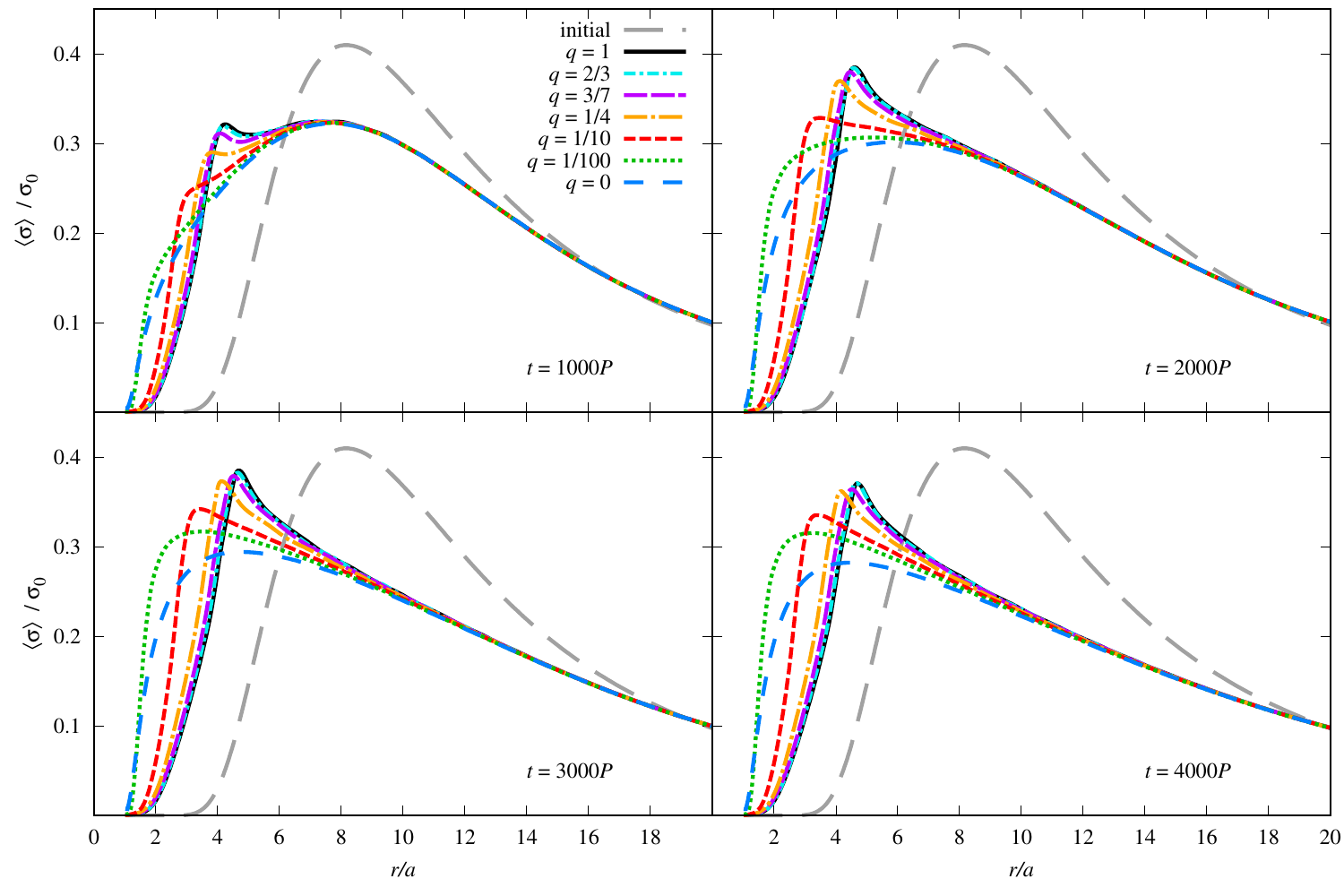}
\caption{Averaged surface mass density for all of the mass ratios summarized in Table \ref{tab:1}, with each panel considering one of the evolution times $t \in \{1000 P, 2000 P, 3000 P, 4000 P\}$. The initial density is also shown in each panel for reference.}
\label{fig:time_sep}
\end{figure}

Figure \ref{fig:not_avg} presents the azimuthally averaged surface mass density $\bar{\sigma}$ for $q \in \{0, 1/100, 1/10, 1\}$ at $t=4000 P$. All cases have smaller density maxima than the initial peak density due to mass loss through the outer boundary. These cases also show varying degrees of periodic structure related to the binary's periodic gravitational potential, which is obviously not present in the single mass ($q=0$) case. Figure \ref{fig:dens_sep} presents the averaged mass density $\langle \sigma \rangle$, calculated using $\Delta=50 P$, as a function of radius for $q \in \{0, 1/100, 1/10, 1/4, 3/7, 2/3, 1 \}$, i.\,e.\, for all of the cases we simulated and summarized in Table \ref{tab:1}. We emphasize that the central part of the system, which is most affected by the periodic binary potential, stabilizes between $3000 P$ and $4000 P$ for all cases. Our conclusions about the governing dynamics and our description of the final gap size are based on simulations as long as $12000 P$, to ensure that the observed quasi-steady state is maintained. The location of the density maximum moves inward as the mass ratio decreases from $1$ to $1/100$, but moves outward again for the single mass case. However, Figure \ref{fig:time_sep}, which presents four different stages of the systems' evolution at $t=1000P$, $t=2000P$, $t=3000P$, and $t=4000P$, shows that the secular trend of an inward migration of the gap continues all the way from the equal mass case through the single mass case.


Following \cite{MacFadyen_2008} and \cite{DOrazio},  we calculate the dynamical torque density
\begin{equation} \label{eqn:d_torque_dens}
\frac{dT_\mathrm{d}}{dr} = -2\pi r \left\langle \sigma         \frac{d\Phi}{d\phi} \right\rangle,
\end{equation}
the integrated dynamical torque
\begin{equation} \label{eqn:d_torque}
    T_\mathrm{d} = \int^r_a \frac{dT_\mathrm{d}}{dr} dr,
\end{equation}
the viscous torque
\begin{equation} \label{eqn:visc_torque}
    T_\nu = 2 \pi r^3 \nu \left\langle \sigma \frac{\partial}{\partial r} \frac{v^\phi}{r} \right\rangle,
\end{equation}
and the viscous torque density $\frac{dT_\nu}{dr}$.

\begin{figure}
\includegraphics[width=0.499\linewidth]{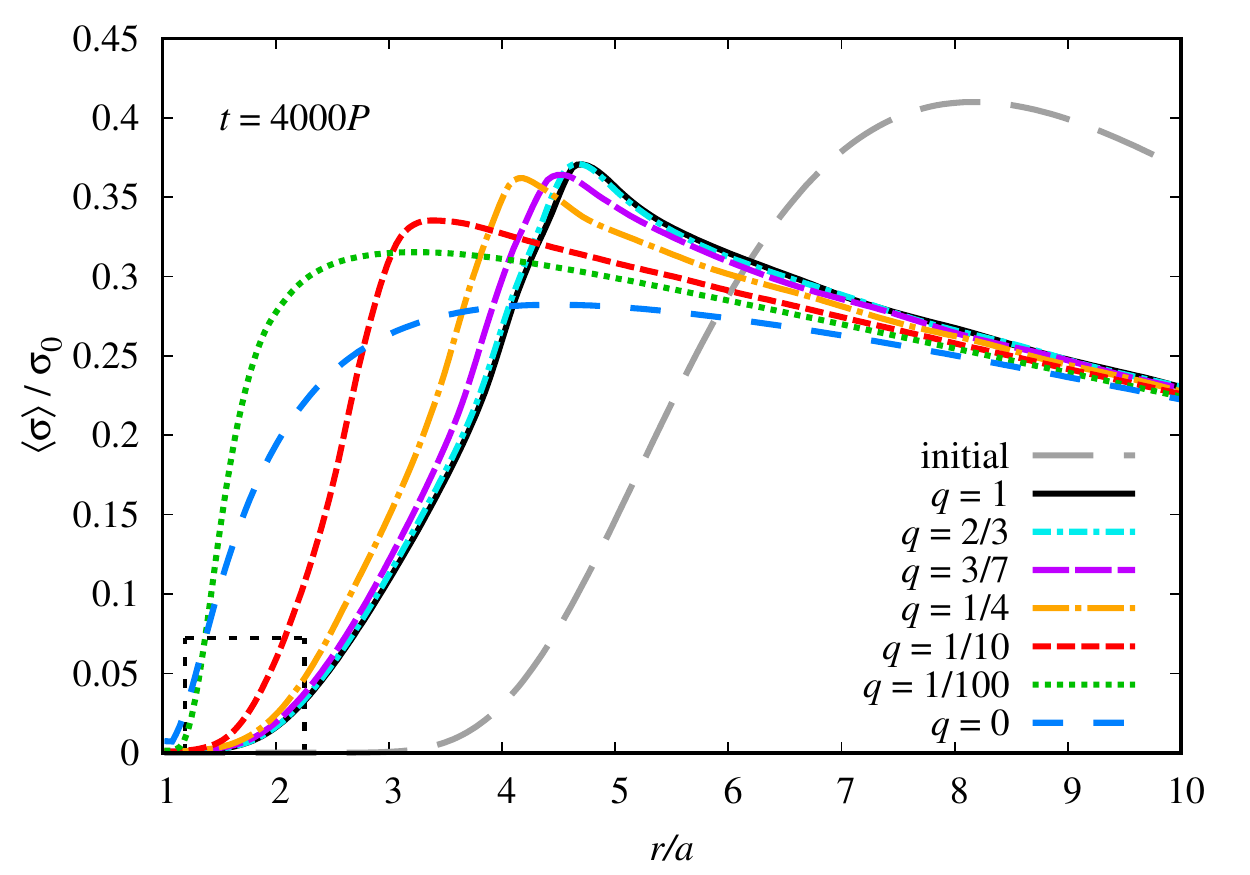}
\includegraphics[width=0.499\linewidth]{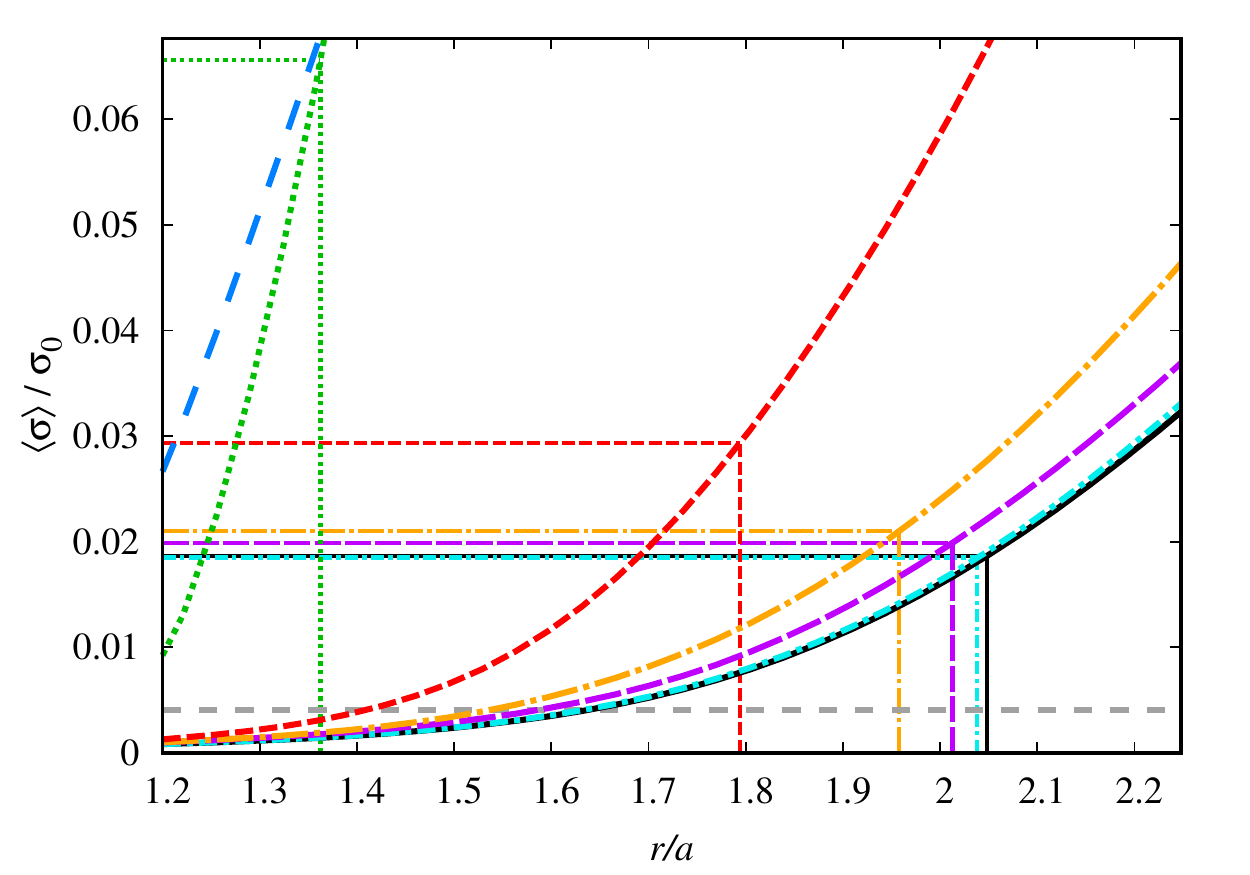}
\caption{Averaged surface mass density at $t=4000 P$ for all cases in Table \ref{tab:1}, along with the initial density. The right panel is a zoom-in of the region inside the dashed box in the left panel. The right panel also shows $r_\mathrm{dT}$ and $\rho(r_\mathrm{dT})$ as we have defined them in the text.}
\label{fig:final}
\end{figure}

\begin{figure}
\includegraphics[width=\linewidth]{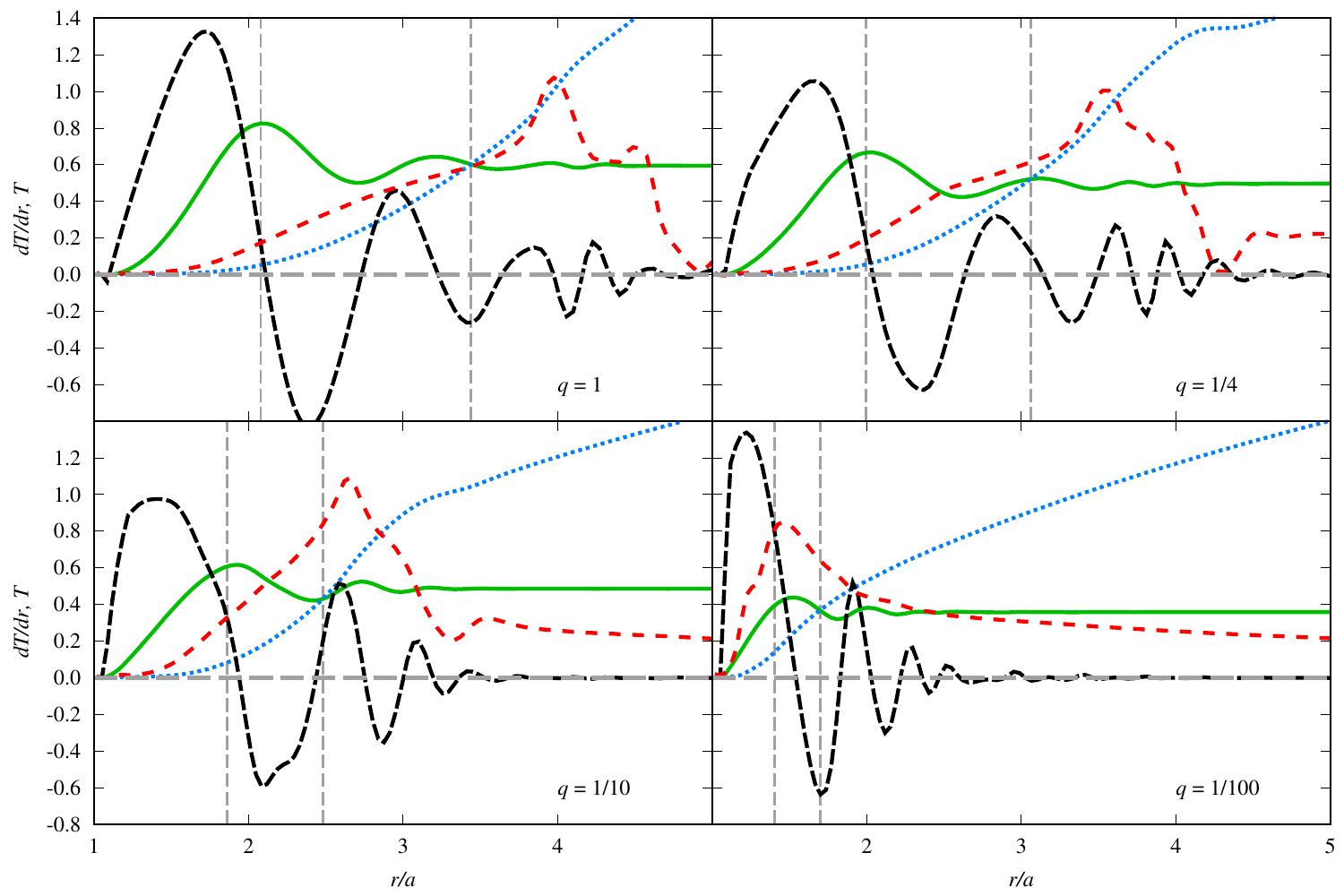}
\caption{Four different types of torques for four different mass ratios. Solid lines correspond to torques, broken lines correspond to torque densities. Red lines correspond to dynamical torques and green lines correspond to viscous torques. Vertical broken lines help to read the intersection radii.}
\label{fig:torque}
\end{figure}

Figure \ref{fig:final} shows the final density profile overall in comparison to the initial data for all of the simulated cases (left panel), as well as a zoom-in on the innermost part of the system (right panel) where all of the potential definitions for the location of a gap occur. Figure \ref{fig:final} specifically illustrates $r_\mathrm{dT}$ --- the radius where the viscous torque density equals the dynamical torque density. 
This concept is further highlighted in Figure \ref{fig:torque}, which shows both types of torque (viscous and dynamical) and their densities as a function of radius. Vertical dashed lines are placed at the radii where the two different types of torque (densities) intersect. 

\begin{figure}
\includegraphics[width=0.49\linewidth]{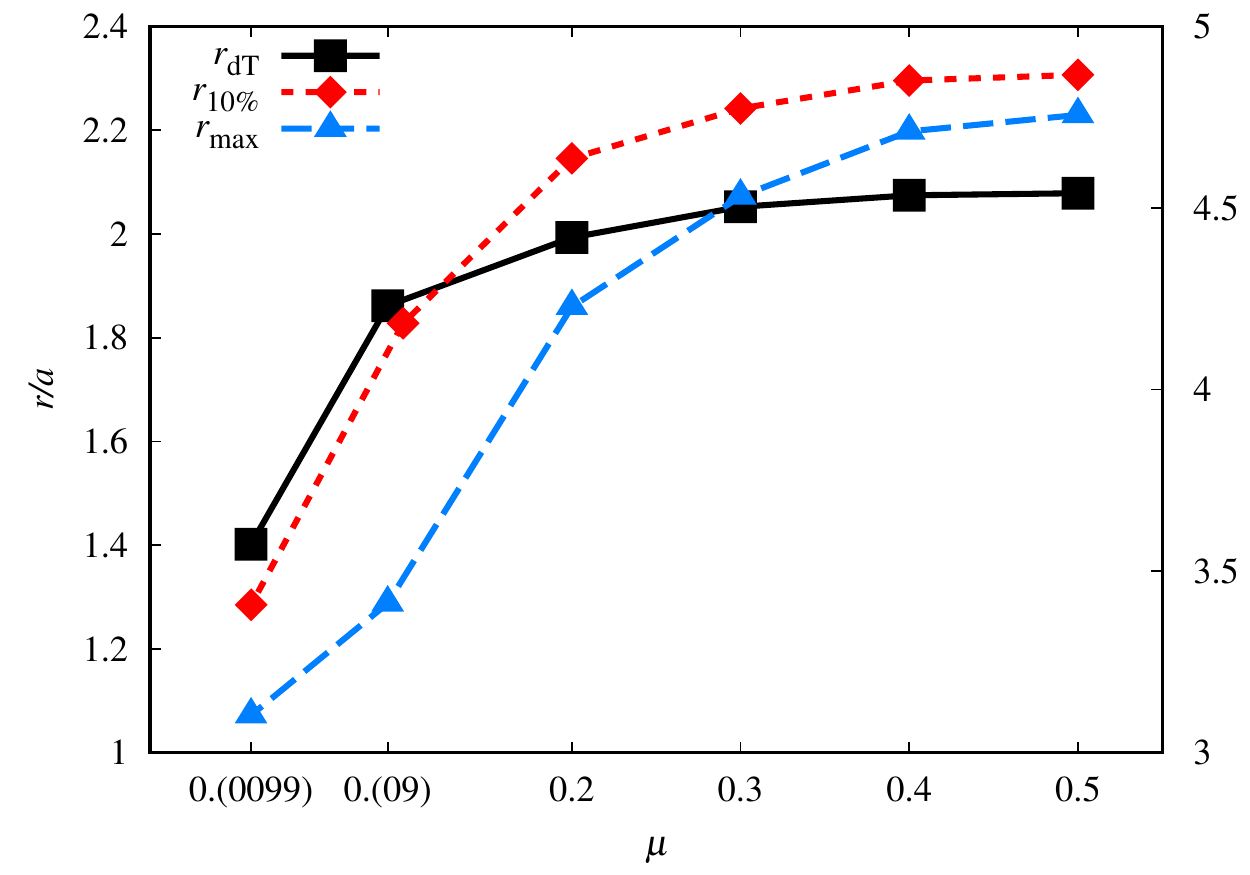}
\includegraphics[width=0.49\linewidth]{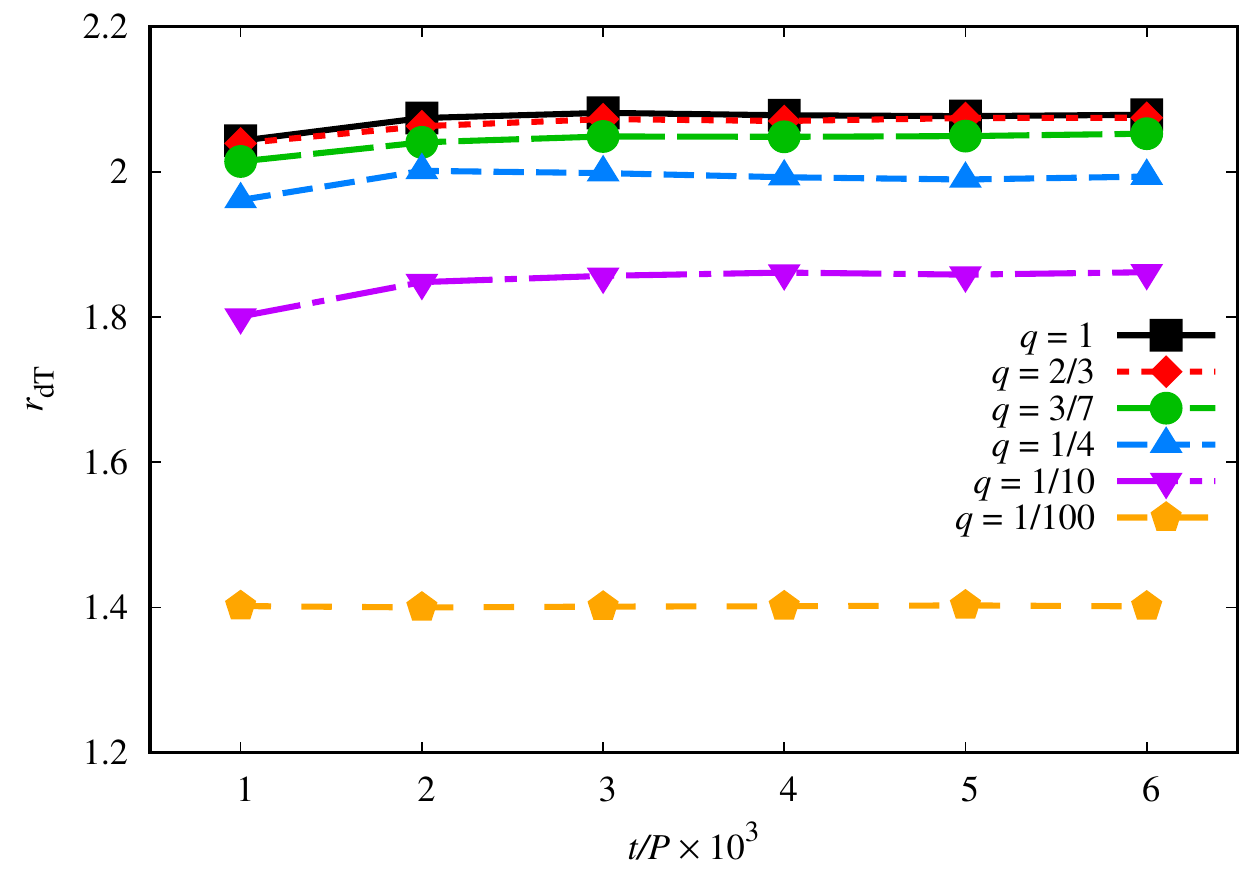}
\caption{Three crucial radii which characterize each configuration as a function of binary mas function $\mu$. The squares denote $r_\mathrm{dT}$ --- the radii where the dynamical torque densities intersect viscous torque densities, the diamonds denote $r_\mathrm{10\%}$ --- the radii where the disk's density reaches $10\%$ of the maximal value. For those two, the scale on the left vertical axis is applied. Finally, triangles denote $r_\mathrm{max}$ --- the location of the density maximums. In this case the right vertical axis is valid. Let us underline that the values of the densities $\sigma (r_\mathrm{dT})$ are different for each configuration. The details are in the Table \ref{tab:1}.}
\label{fig:points}
\end{figure}

In Table \ref{tab:1}, in addition to the mass ratio $q$ and the secondary mass ratio $\mu$, we have listed the following system characteristics for all of our simulated cases: the radial location of the maximum density, $r_\mathrm{max}$, and the maximum density value $\rho(r_\mathrm{max})$; the radius where the torque densities balance, $r_\mathrm{dT}$, and corresponding torque density value, $\rho(r_\mathrm{dT})$; and $r_\mathrm{10\%}$, the radius where the density inside the cavity falls to $10\%$ of the final density maximum. This last quantity is introduced as an additional intuitive way to define a gap. The specific choice of $10\%$ is essentially arbitrary. We note that $r_\mathrm{10\%}$ is found by interpolation, and is not therefore restricted to occur at a cell boundary. To show their general dependence on the mass ratio, all of these radii are plotted in the left panel of Figure \ref{fig:points}. The right panel of Figure \ref{fig:points} shows the stability of the $r_\mathrm{dT}$ with time; its value is established very early, by the time $t$ reaches $1000 P$ -- $2000 P$, while the other referenced radii (especially $r_\mathrm{max}$) are still actively evolving towards their quasi-steady states.

\subsection{Azimuthal mode analysis}
\label{sec:mode}
\begin{figure}
\includegraphics[width=\linewidth]{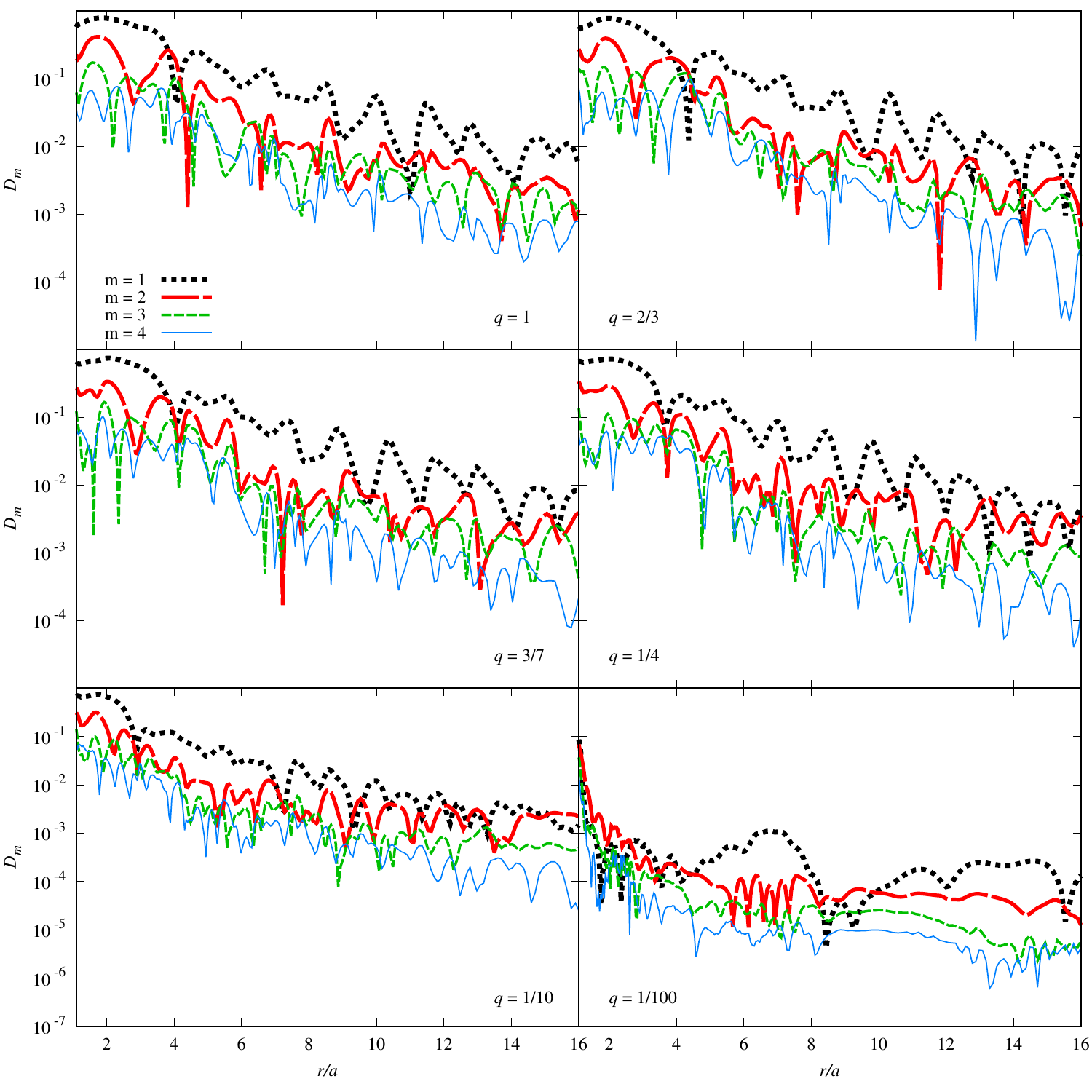}
\caption{Azimuthal modes decomposition of the density at $t=4000P$ for all of our simulated cases.}
\label{fig:modes}
\end{figure}

An important aim of this study is to better understand the nature of angular momentum transport in circumbinary disks as a way to interpret astrophysical disk observations and what they can tell us about the underlying dynamics. A frequent primary assumption in previous analytic studies of these disks has been the perturbative treatment of azimuthal symmetry breaking due to the binary potential. In order to provide a measure of accuracy for such a perturbative treatment, and to better assess the importance of the \textit{resonant} transfer of angular momentum, we compute the corresponding azimuthal breakdown of the density that indicates the presence or absence of a perturbative effect. 
In Figure \ref{fig:modes}, we present the decomposition of the density distribution into azimuthal harmonic modes for all of our simulated cases. This decomposition is given by
\begin{equation}
D_m = \frac{1}{2\pi^2} \int_0^{2\pi} d\phi \int_0^{2\pi} d(\Omega_\mathrm{b} t) \rho e^{i m (\phi - \Omega_\mathrm{b} t)},
\end{equation}
where $\Omega_\mathrm{b}$ is the binary's orbital angular frequency. In the perturbative picture presented in previous studies \citep{AL94, GT1980ApJ}, these modes are driven by the force associated with the corresponding harmonic mode of the gravitational potential. Since each harmonic mode represents a contribution from a higher multipole moment, we expect the strength of the driving force to weaken with higher mode number $m$, and for odd harmonic modes to weaken as we approach the equal mass due to the absence of odd-$m$ potential modes in that case.
However, as illustrated in Figure \ref{fig:modes}, we do not see an unambiguous decay with increasing mode number beyond $m=1$ at any length scale across the next three $m$ modes, with the possible exception of the $q=1/10$ and $1/100$ cases at large radii. We also note the dominance of the $m = 1$ mode everywhere except the $q=1/100$ case at small radii, particularly its dominance for the equal mass case, where the contribution from any odd $m$ mode of the gravitational potential vanishes. This suggests a departure from a linear coupling of the potential and density modes in the dynamics. Such a coupling would be responsible for the transport of angular momentum and the formation of a circumbinary gap in the resonant torquing picture, so its absence strongly suggests the dynamical irrelevance of resonant torquing in this scenario. We will discuss the consequences of this and make suggestions for a better analytic treatment of circumbinary disk dynamics in the next section.

\section{Analytical Calculations} \label{sec:sid}

\subsection{Orbital Stability} \label{subsec:stable}

In order to reconcile the results from the hydrodynamics simulations with analytic theory, and thereby obtain a coherent understanding of circumbinary disk behavior, it is important to look at the stability of the approximate solutions derived in Sec. \ref{sec:preliminary_sid}. We note that in order to translate from the hydrodynamic picture to the test particle picture, the disk is considered to be a collection of fluid elements which each occupy an infinitesimal area centered about a hypothetical test particle. Then, the orbit of the test particle acts as an approximation for the motion of this fluid element in the absence of viscosity. In that case, the exponential blow-up of small perturbations about the test particle's orbit, especially over timescales comparable to or shorter than the binary orbital period, provides an important insight into the effect of the binary potential on the stability of the fluid elements themselves. In other words, instability in the test particle picture suggests that the corresponding fluid elements in the hydrodynamic picture will either be expelled to larger radii or accreted by the binary.

The physical quantity relevant to orbital stability is the Lyapunov exponent. For a Hamiltonian system with phase space variables $\eta$, the equation of motion is given in the symplectic form as (\textit{cf}. Eq. 8.39 of \cite{Goldstein})
\begin{equation}
    \dot{\eta}_i = \bar{I}_{ij}\frac{\partial\mathcal{H}}{\partial\eta_j}.
\end{equation}
Where, $\bar{I}_{ij}$ maps the derivatives of the Hamiltonian to the appropriate time derivative with the appropriate sign. Let us assume there exists a particular solution, $\eta^{(0)}$, to the above equations of motion. Since a particular solution can be entirely specified by a choice of initial conditions, the stability of that solution can be defined as the extent to which a new solution, $\eta^{(1)}$, with approximately similar initial conditions, departs from the original solution. To quantify this, let us assume that $\eta^{(1)} = \eta^{(0)} + \delta \eta$. Here, $\delta \eta$ represents the difference between the two solutions with $\delta \eta(t = 0)$ being infinitesimally small. We can then linearize the above equations of motion to get
\begin{equation}
    \delta\dot{\eta}_i = \bar{I}_{ij}\frac{\partial^2\mathcal{H}}{\partial\eta_j\partial\eta_k}\bigg\rvert_{\eta^{(0)}}\delta\eta_k.    
\end{equation}
Thus, the departure of the new solution due to a small initial perturbation $\delta\eta(0)$ is given by the matrix equation
\begin{eqnarray}
    \delta\dot{\mathbf{\eta}} & = & \mathds{K}\delta\mathbf{\eta}, \\
    K_{ij} & = & \bar{I}_{im}\frac{\partial^2\mathcal{H}}{\partial\eta_m\partial\eta_j}\bigg\rvert_{\eta^{(0)}}.
\end{eqnarray}
The above system of differential equations has a general solution given by a superposition of the eigenvectors of $\mathds{K}$ which evolve exponentially at a rate given by the corresponding eigenvalues, $\lambda$, which are known as the Lyapunov exponents. Since $\mathds{K}$ will in general have complex values, the real part, $\mathfrak{Re}(\lambda)$, gives the timescale over which the perturbations decay (if $\mathfrak{Re}(\lambda) < 0$), or blow up (if $\mathfrak{Re}(\lambda) > 0$); the imaginary part, $\mathfrak{Im}(\lambda)$, gives the period over which the perturbations oscillate. We note that the eigenvectors of the stability matrix represent the direction in phase space of the initial perturbations associated with each of these exponents. A particular orbit may be metastable, which indicates that one set of perturbations tends to decay while an orthogonal set may blow up.
It is useful to study the stability of particular ``families'' of orbits, as we do with the epicyclic orbits in this study.

\subsubsection{Stability Matrix for Epicyclic Orbits}

Using the tools from Sec.\,\ref{subsec:stable}, we now study the stability of circumbinary orbits under the epicyclic approximation. Since we will solve the equations perturbatively, we approximate the stability matrix to the same order as the linearized solution for consistency. To do so, we introduce a bookkeeping parameter, $\epsilon$, such that all the non-axisymmetric potential terms and their corresponding perturbations on the circular orbits will enter at leading order. Thus, the Hamiltonian presented in Sec.\,\ref{sec:preliminary_sid} is given by
\begin{equation}
    \mathcal{H} = \frac{p_r^2}{2} + \frac{p_\varphi^2}{2r^2} - p_\varphi - \frac{1}{r} + \epsilon\Phi_m(r)\cos(m\varphi).    
\end{equation}
For the above Hamiltonian, the stability matrix, $\mathds{K}$, takes the form
\begin{equation} \label{eq:K}
    \mathds{K} = \left(
\begin{array}{cccc}
 0 & 0 & 1 & 0 \\
 -\frac{2 l}{r^3} & 0 & 0 & \frac{1}{r^2} \\
 -\frac{3 l^2 }{r^4} + \frac{2}{r^3} - \epsilon \Phi''_m   \cos (m \varphi )  & \epsilon \Phi'_m \cos (m \varphi ) & 0 & \frac{2 l}{r^3}\\
 m \epsilon \Phi'_m \sin (m \varphi) & m^2 \epsilon \Phi_m \cos (m \varphi) & 0 & 0\\
\end{array}
\right).
\end{equation}
We evaluate Eq.\,\eqref{eq:K} for the approximate orbital solution
\begin{eqnarray}
r &=& r_0 + \epsilon r_1,\\
p &=& \epsilon\frac{r_1}{\omega},\\
l &=& \sqrt{r_0} + \epsilon l_1.
\end{eqnarray}
The first order solutions to the above expressions can be found in Eqs.\,\eqref{epicycle::L} and \eqref{epicycle::R}. We note again that we do not need a first order solution for $\varphi$ since it enters the matrix at $\mathcal{O}(\epsilon)$. Thus, we can use
\begin{eqnarray}
    \varphi = \omega t, \\ \nonumber
    \omega = \sqrt{\frac{1}{r_{0}^{3}}} - 1.
\end{eqnarray}
Upon substitution, we Taylor expand the elements of the matrix and set $\epsilon = 1$. The stability matrix can then be split as
\begin{equation}\label{epicycle:K}
\mathds{K} = \mathds{K}_0 + \mathds{\delta K},
\end{equation}
where $\mathds{K}_0$ comes from the Keplerian background motion and $\delta \mathds{K}$ contains all the perturbations. $\mathds{K}_0$ is given by
\begin{equation}\label{kepler::K}
    \mathds{K}_0 = \left(
    \begin{array}{cccc}
        0 & 0 & 1 & 0 \\
        -\frac{2 l_0}{r_0^3} & 0 & 0 & \frac{1}{r_0^2} \\
        -\frac{1}{r_0^3} & 0 & 0 & \frac{2 l_0}{r_0^{3}} \\
        0 & 0 & 0 & 0 \\
    \end{array}
    \right),
\end{equation}
and the perturbative matrix $\delta \mathds{K}$ is
\begin{equation}
    \delta\mathds{K} = \left(
    \begin{array}{cccc}
        0 & 0 & 0 & 0 \\
        -2\left(\frac{ l_1}{r_0^3} + \frac{3l_0r_1}{r_0^4}\right) & 0 & 0 & -2\frac{r_1}{r_0^3} \\
        6\left(\frac{r_1}{r_0} - \frac{l_0l_1}{r_0^4}\right) - \Phi_m''\cos(m\phi) & m\Phi_m'\sin(m\phi) & 0 & 2\left(\frac{l_1}{r_0^{3}} - \frac{3l_0r_1}{r_0^4}\right) \\
        m\Phi_m'\sin(m\phi) & m^2\Phi_m\cos(m\phi) & 0 & 0 \\
    \end{array}
    \right).
\end{equation}

\subsubsection{Lyapunov Exponents for Epicyclic Orbits} \label{sec:Lyapunov_Exponents}

The computation of closed-form eigenvalues for the stability matrix given in Eq.\,\eqref{epicycle:K} at linear order is complicated by the fact that regular matrix perturbation theory does not hold for degenerate matrices. Since the Keplerian stability matrix, Eq.\,\eqref{kepler::K}, has a doubly-degenerate, trivial eigenvalue of zero, it is not invertible, and therefore we cannot proceed to solve the eigenvalue problem analytically; we therefore use the Python package {\sc SciPy} \citep{SciPy} to numerically compute the eigenvalues.

We compute the eigenvalues for a family of epicyclic orbits around the background radius $r_0$ in the range $r_0 \in [1.1,3.0]$, as this covers the relevant resonances and the typical gap scale, which we will define below. For the binary mass ratio, we chose the range $\mu \in [0.01,0.5]$, as this is the range probed by the numerical studies.

\begin{figure}
\begin{center}
\includegraphics[width=0.6\linewidth]{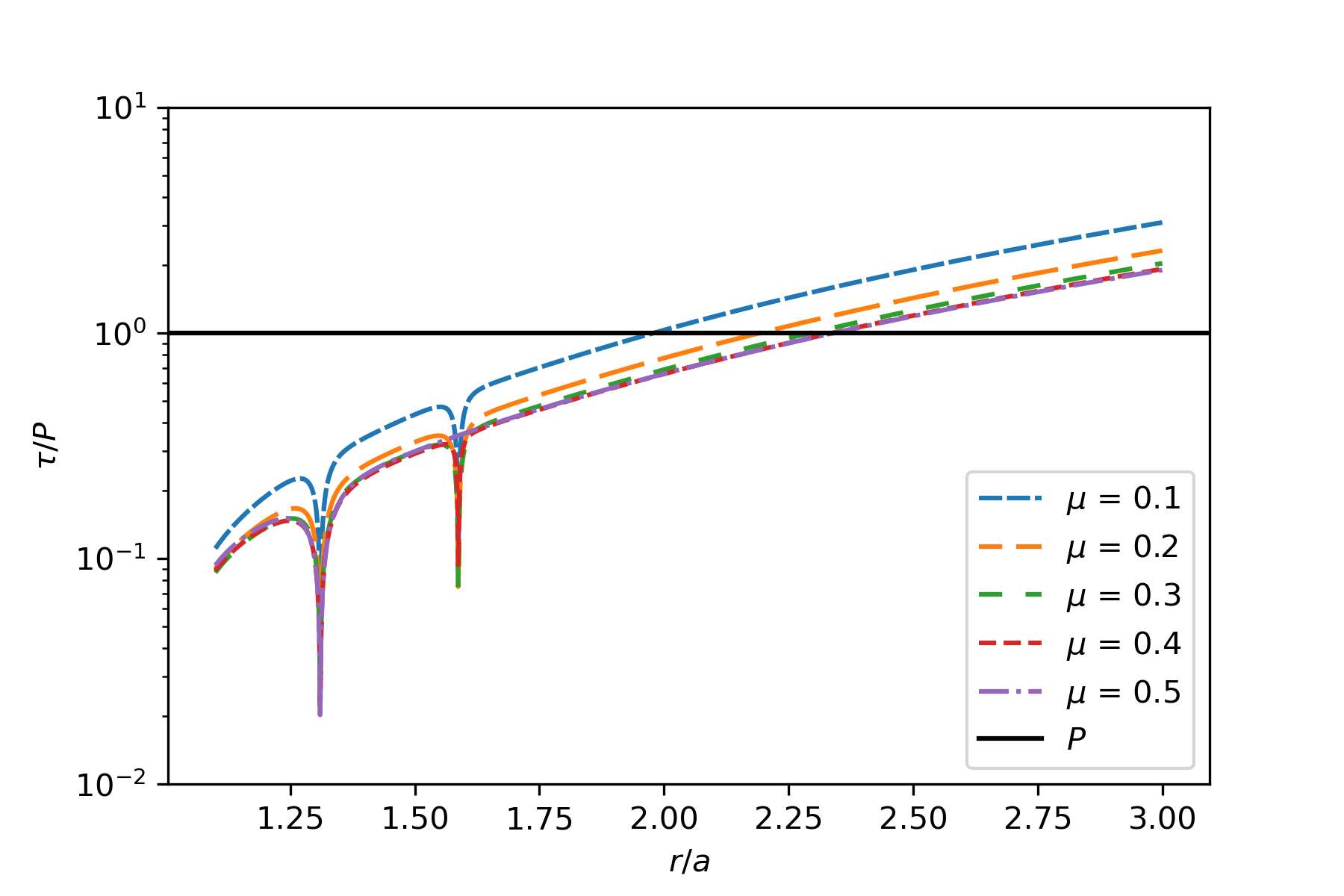}
\caption{The dependence of the quantity $\tau$ defined in Eq.\,\eqref{eqn:tau} on the radius. The black line represents the binary period and $\tau$ is normalized by the binary period for convenience.}
\label{fig:s1}
\end{center}
\end{figure}

We separate the real part of the Lyapunov exponents, and search for the largest positive value at each radius. Since the real part is the inverse timescale over which instabilities propagate, the largest value corresponds to the shortest timescale, which we will refer to as the Lyapunov timescale. To define a gap size $r_\mathrm{L}$ based on the Lyapunov timescale $\tau$, we find the radius that satisfies
\begin{equation}\label{eqn:tau}
    \tau(r_\mathrm{L}) \equiv \left[\mathfrak{Re}\left(\lambda(r_\mathrm{L})\right)_{\mathrm{max}}\right]^{-1} = P,
\end{equation}
where, $\lambda(r_\mathrm{L})$ denotes the Lyapunov exponent at $r_\mathrm{L}$, ``max'' indicates that we are choosing the Lyapunov exponent with the largest real part, and $P=1$ in our units.
In Figure \ref{fig:s1}, we present the Lyapunov timescales for a range of systems. We note that there are regions where the timescale $\tau$ drops drastically. We confirm that these are the locations of the Lindblad Resonances where $\tau \rightarrow 0$ or, more importantly, $\lambda \rightarrow \infty$. This effect only demonstrates a \textit{complete} breakdown of the epicyclic approximation. However, as it can be seen from the rest of the graph, the epicyclic approximation is unstable for radii larger than the resonances. \cite{BinneyTremaine} provide a method for regularizing, i.e, removing the infinite behavior at the resonances. We did not perform this regularization as it does not affect the outcome of the gap sizes, which occur at larger radii than the resonances.
The rationale behind the choice $\tau = P$ in Eq.\eqref{eqn:tau} is that, while the binary still induces instabilities at larger radii, the Lyapunov timescale increases with radius far faster than the orbital period. 

We finally emphasize that this definition is independent of the nature of the fluid characterization. The study of test particles indicates that instabilities propagate over timescales shorter than an orbit, and therefore far shorter than the characteristic viscous timescale as believed in \cite{GT1980ApJ}. Therefore, if this is a viable explanation for the formation and maintenance of a circumbinary gap, then it should be independent of the viscous prescription as noted in \cite{DOrazio}. 

\subsection{Resonant Torquing Picture}
\label{sec:Resonant Torquing Picture}
Prior analytical studies of circumbinary accretion disks have arrived at a criterion for the opening of a circumbinary gap from hydrodynamic considerations (\textit{cf}.\,\cite{AL94} and \cite{GT1980ApJ}). Instead of studying the stability of orbits due to the gravitational potential of the binary, those works have explored the behavior of the primitive fluid variables that evolve under the influence of the binary. In that case, the WKB approximation to the equations of fluid mechanics can be used to study the behavior of the primitive variables over timescales that are larger than the binary timescale but still comparable to or shorter than the viscous timescale. 

While the two pictures of test particles and hydrodynamic flows can be related by associating particle orbits with steady streamlines of the gas, they differ in the physical mechanism that is invoked to explain the opening and maintenance of a circumbinary gap. In our analysis, the opening and maintenance of the gap is contingent on the divergence of an infinitesimal element over short timescales due to the tidal nature of the binary potential; in the fluid picture, the gap is contingent on the ability of the Lindblad resonances associated with the binary potential to transfer angular momentum outward through the disk faster than viscous damping forces can dissipate it. The mechanisms primarily differ over the role of resonances in the formation of the gap and the timescales over which the binary potential maintains the gap. In this subsection, we will recap some of the quantities derived in \cite{GT1980ApJ}, proceed to define the criterion for a resonance to be gap-opening, and compute gap sizes from this definition.

The gap opening timescale is defined as the time it takes for a torque at a Lindblad resonance to deposit the angular momentum needed to open a gap in the form of an infinitesimal ring of width $\Delta r$ at a radius $r$. To quantify this, we first to define the amount of angular momentum, $\Delta H$, required to open such a gap in a disk with surface density $\sigma$ as
\begin{equation}
\Delta H = \sigma \Omega \left(r\Delta r\right)^2.
\end{equation}
If $T_{\mathrm{LR}}$ is the torque delivered by a Lindblad resonance, then the gap opening timescale is given by
\begin{equation}\label{GT1980::torque}
\Delta t_{\mathrm{open}} = \frac{\Delta H}{T_{\mathrm{LR}}}.
\end{equation}
\cite{GT1980ApJ} also give the expression for the torque at the $m^{\mathrm{th}}$ Lindblad resonance as
\begin{equation}
T_m = -m\pi^2\left[\sigma \left(\frac{d D}{d \ln r}\right)^{-1} \left|\Psi_m\right|^2 \right],
\end{equation}
where $\Psi_{m}$ is defined as
\begin{equation}
\Psi_{m} = \frac{d \Phi_m}{d \ln r} - 2m\Phi_{m},
\end{equation}
and $D$ is defined in terms of the epicyclic frequency, $\kappa$, and the orbital frequency, $\omega$, as $D \equiv \kappa^2 - m^2\left(\omega-1\right)^2$. For the Keplerian background ($\kappa = \omega = r^{-3/2}$) considered below, $\frac{d D}{d \ln r}$ evaluated at the Lindblad resonance is then given by
\begin{equation}
     \frac{d D}{d \ln r}\bigg\rvert_{\mathrm{LR}} = - \frac{3m^2}{m+1}
\end{equation}

To complete this picture of gap maintenance due to Lindblad resonances, we also need the gap closing timescale, which is defined as the time it will take for viscous damping forces to close a ring-shaped gap of similarly infinitesimal radial size $\Delta r$ at a radius $r$,

\begin{equation}
\Delta t_{\rm close} = \frac{(\Delta r)^2}{\nu}
\end{equation}
Where $\nu$ is the coefficient of viscosity for the accretion disk. For the $\alpha$-disk prescription, again assuming a Keplerian background, one has from Eq.\,\eqref{alpha_type}
\begin{equation}
\nu = \frac{\alpha \chi^2}{r\Omega} 
\end{equation}
where $\alpha$ and $\chi$ are defined in Sec.\,\ref{sec:preliminary_mp}.

We therefore arrive at the first gap opening criterion for this picture, which states that a resonance is considered gap opening if the gap closing timescale exceeds the gap opening timescale. The gap opening resonance is the outermost Lindblad resonance where the criterion $\Delta t_\mathrm{close} > \Delta t_\mathrm{open}$ is met, or equivalently, where the parameter $\zeta_{\Delta t}$ given by
\begin{equation}
    \zeta_{\Delta t} = \frac{\Delta t_\mathrm{close}}{\Delta t_\mathrm{open}} = \frac{1}{\alpha \chi^2}\frac{\bar{T}_m}{r}
\end{equation}
exceeds unity, where the term $\bar{T}_m$ refers to the torque as defined in Eq.\,\eqref{GT1980::torque}, normalized by the surface density.

A second gap opening criterion which is used in \cite{AL94} and \cite{2015MNRAS.452.2396M} considers the balance between the viscous and resonant torques instead of their associated timescales. It requires that the resonant torque, which clears the gap, be stronger than the viscous torque as defined in Eq.\,\eqref{eqn:visc_torque} that brings fluid in to fill the gap. These two criteria differ by a factor of $3\pi$. We define in this case a corresponding parameter, $\zeta_T$, such that
\begin{equation}
    \zeta_{T} = \frac{T_m}{T_\nu} = \frac{1}{3\pi\alpha \chi^2}\frac{\bar{T}_m}{r} = \frac{\zeta_{\Delta t}}{3\pi}. 
\end{equation}
We note that, unlike the numerical computation of the torque in Eq.\,\eqref{eqn:visc_torque}, we do not need azimuthal averaging, since the background densities used to derive these expressions are already spherically symmetric.
While the parameter $\zeta_{T}$ provides a stronger check of gap opening due to the factor $3\pi$, we note that both criteria ultimately yield the same results for the gap opening resonance. 
In order to obtain a measure of gap size, we must refine the range of influence of the gap opening resonance, since the torque is not, in reality, deposited at the discrete resonance radius, but rather it is spread over some neighborhood surrounding the resonance. In \cite{AL94}, this spread is addressed by considering the extent to which intersecting fluid streamlines bundle around the resonances. We take this effect into consideration by turning, once again, to the theory of epicycles. We define the gap size to be the outermost radius where the amplitude of epicycles allows for the orbit to intersect the resonance location. If we label the amplitude of the epicyclic solution $A \equiv A(r_0)$ by its background radius $r_0$, then we can define the gap size at $r_T$ to obey the condition

\begin{equation}
    A(r_T) = |r_T - r_\zeta|,
\end{equation}
where $r_\zeta$ is the location of the gap opening resonance. Since this equation is non-linear, we invert it to find $r_T$ using a simple bisection solver since the form of the amplitude of epicycles is strictly decreasing from the location of the resonance. In order to better resolve the solution to the above equation, we choose to perform the bisection search on the logarithmic version of the same condition, that is

\begin{figure}
\begin{center}
\includegraphics[width=0.6\linewidth]{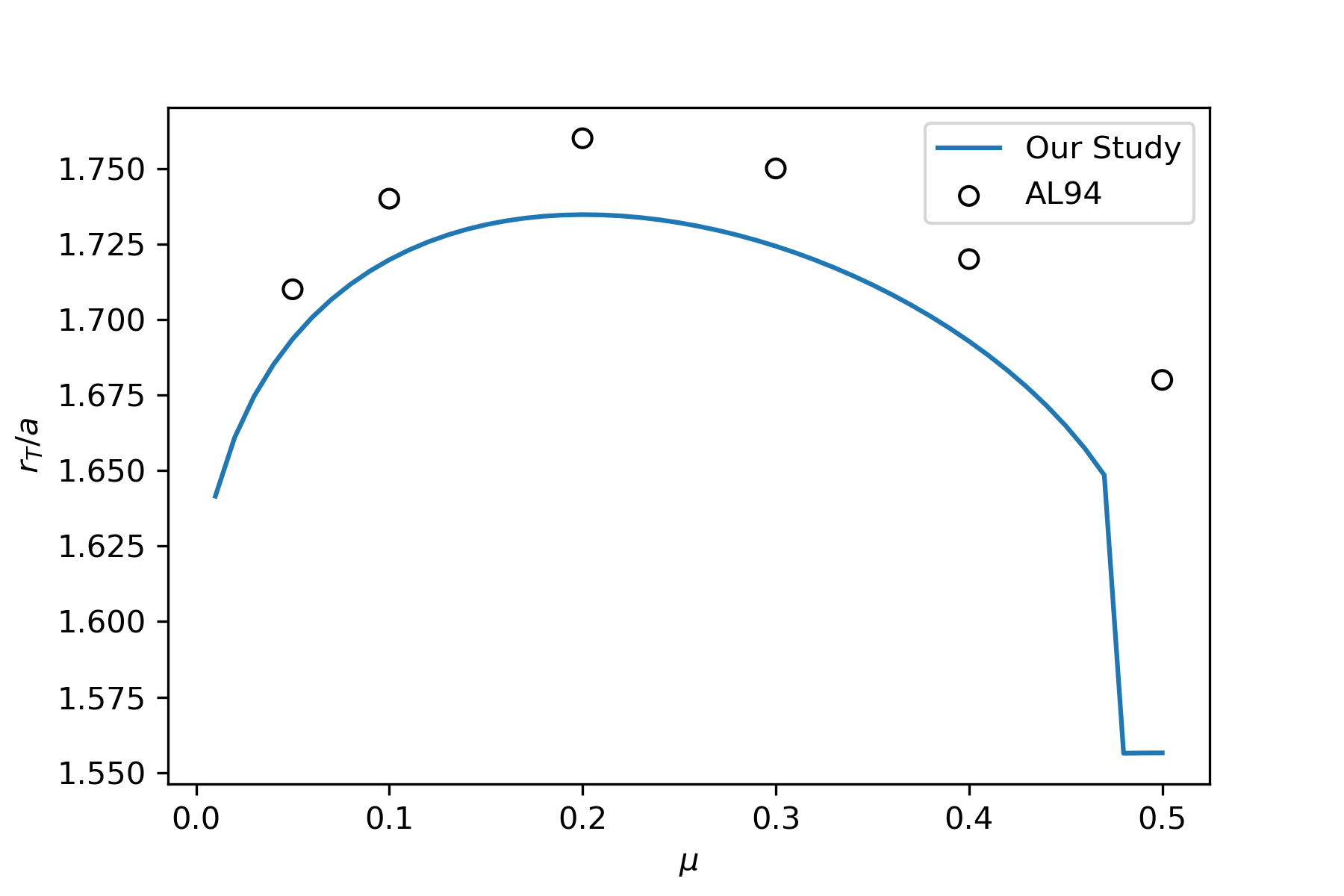}
\caption{Comparison of the dependence of $r_T$ from Eq.\eqref{eq:rT} and the gap sizes from Table 1 of \cite{AL94} over mass ratio $\mu$. We find that there is an overall agreement of the trend against mass ratios except for the offset at $\mu \approx 0.5$. The torque from the $m=1$ resonance abruptly drops at $\mu \approx 0.5$ leading to an immediate dominance of the $m=2$ resonance.}
\label{fig:s3}
\end{center}
\end{figure}

\begin{equation}\label{eq:rT}
    f(r_T) \equiv \log\left(\frac{A(r_T)}{|r_T - r_\zeta|}\right) = 0.
\end{equation}
We present the results of the above gap size computation along with results from \cite{AL94} to show that they are consistent.

\subsection{Comparison of Analytical and Numerical Results}

Having arrived at multiple definitions of the gap size, we note that there is a stark contrast in the physics relevant to maintaining a circumbinary gap between the two analytical pictures. In the test particle case, the gap is maintained by instabilities propagated by the binary potential over timescales comparable to the binary period, while in the resonant torque case, the gap is maintained by viscous transport of angular momentum deposited at resonances over the viscous timescale. Additionally, the test particle picture would indicate that the gap size should not depend on the viscosity parameters, whereas the resonant torque picture would require a strong correlation between gap sizes and viscosity parameters.

It is also evident that the two pictures predict different trends in the variation of gap size over different mass ratios. In the test particle case, gap sizes increase monotonically with increasing mass ratio. In contrast, in the resonant torque scenario, the gap size initially increases, reaches a maximum, and then decreases with increasing mass ratio, with a more abrupt decrease very close to the equal mass case where the transition from the $m=1$ to the $m=2$ resonance occurs. Thus, it is crucial to look to the numerical simulations of the disk-binary system to arrive at a conclusion as to which physics is actually responsible for creating and maintaining gaps in circumbinary disks.

We emphasize that we wish to compare the trends predicted in these two pictures, rather than the exact values for the gap size at any given mass ratio, since a realistic circumbinary gap is just a diffuse region and does not have a hard boundary. We therefore normalize the results for the several radii that we have defined throughout this work, to clearly identify the trends they make when we overlay them graphically. To that end, we define the quantities $\bar{r}_\mathrm{X}$ such that,
\begin{equation}
    \bar{r}_\mathrm{X} = \frac{r_\mathrm{L}}{r_\mathrm{X}}\bigg\rvert_{\mu = 0.5} r_\mathrm{X},
\end{equation}
where the subscript $\mathrm{X}$ denotes the subscript of the relevant radius from the numerical or analytical computations, e.g., $r_\mathrm{10\%}$ as defined in Sec.\,\ref{sec:nr}. The vertical line denotes that the values are taken at the mass ratio $\mu = 0.5$, so that the gap size as defined by the instability timescale and the numerical/analytical measure of gap size are normalized to agree for equal masses.

\begin{figure} 
\begin{center}
\includegraphics[width=0.6\linewidth]{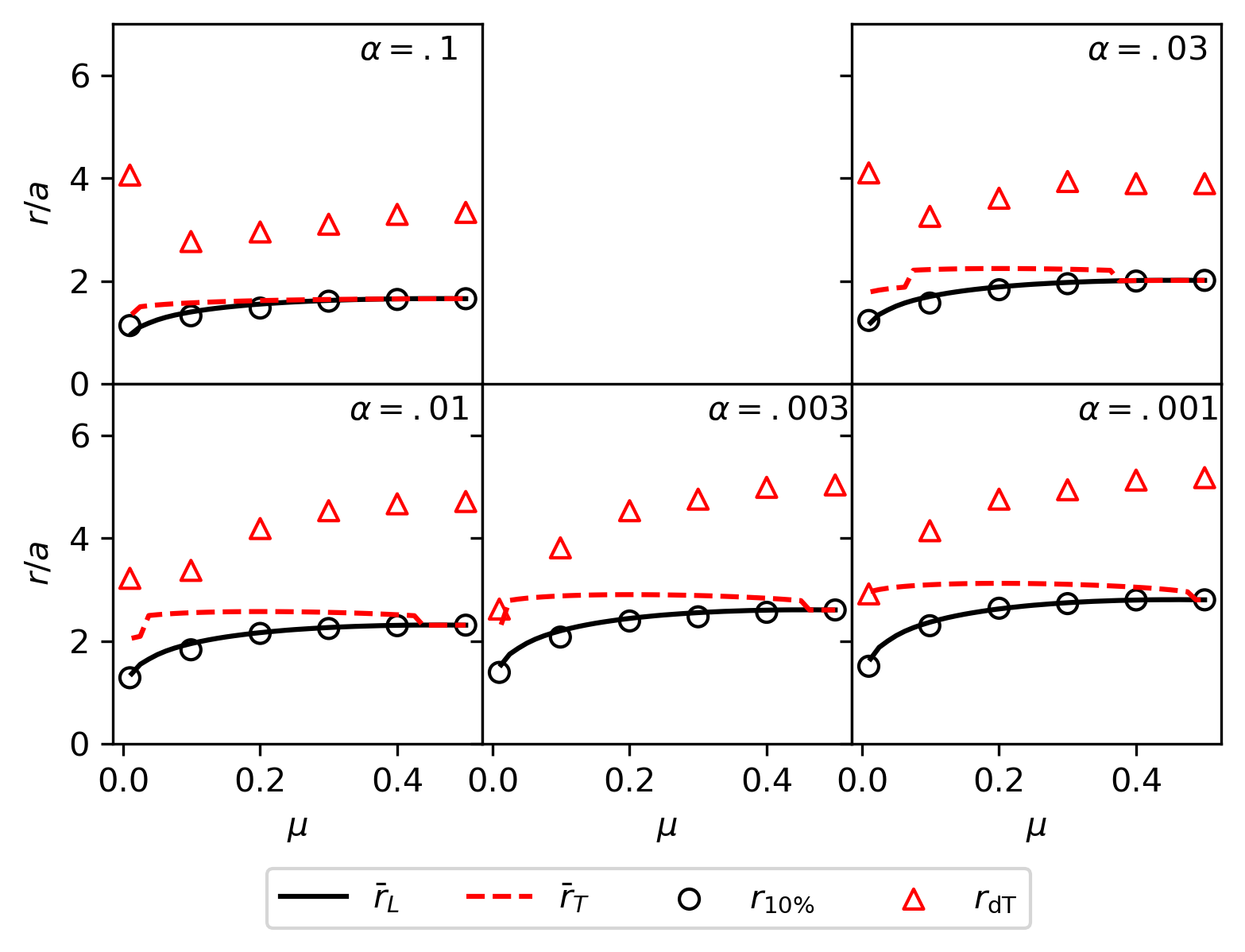}
\caption{ Comparison of numerical results of $r_{10\%}$ and $r_{\mathrm{dT}}$ in Table.\ref{tab:1}, and $r_L$ and $r_T$ computed in Eqs.\,(\ref{eqn:tau},\ref{eq:rT}). We notice that the trends for the numerical radii agree better with the gap size as defined by the instability timescale as opposed to the resonant torque balance criteria for the simulated range of $\alpha$.}
\label{fig:compare}
\end{center}
\end{figure}

We plot the results of this comparison in Figure \ref{fig:compare}, which is the central result of this work. We note the clear agreement in the trend of the numerical results with the gap size as predicted by the instability timescale in the test particle picture, indicating that the physics responsible for the gap truncation and maintenance is the instabilities that are rapidly propagated at sub-binary timescales. 

The disagreement of the resonant torque picture with the numerical results in Figure \ref{fig:compare} reinforces the disagreement we saw in the azimuthal variation of the density distributions in Sec.\,\ref{sec:mode}. In the analytical picture, we expect that each successive multi-polar contribution from the binary potential should affect the disk dynamics ever more weakly, due to the ever larger inverse power-law drop offs of those higher order contributions. Additionally, as we approach the equal mass case, there is a symmetry which ensures that odd multipole moments have zero contribution to the potential, and therefore there should be no density perturbations excited at those harmonics in the resonant torque scenario.

However, the numerical results show a stark contrast from this description. The magnitudes of azimuthal modes for all harmonics are within 2 orders of magnitude of each other, with odd harmonics never being suppressed to any degree as we approach the equal mass case. We note especially in the equal mass case that the $m = 1$ azimuthal perturbations dominate by almost an order of magnitude over the $m = 2$ mode that should be the leading order multipole moment in the analytic theory. Additionally, we note that the strength of the higher azimuthal modes does not decay, which indicates strongly that the different azimuthal modes are being propagated non-linearly.

These results, and the ability of test-particle orbit instabilities to explain the numerically observed behavior of the circumbinary gap, lead us to the conclusion that angular momentum transport in circumbinary disks is not well explained by a competition of dynamical and viscous torques. Specifically, the location of the circumbinary gap does not correlate with the viscosity of the disk. Additionally, we do not observe a drop-off in the circumbinary gap size as we approach equal mass ratios. Azimuthal variations in the density profile are also not modulated in any way by the multipole moments that excite them in the perturbative picture, be it the radial drop-off in individual cases or the weakening of odd multipole moments as we approach the symmetrical equal-mass case.

\section{Conclusions}
\label{sec:conc}

We have provided a comprehensive study of the effect of a binary on a surrounding circumbinary disk through both analytical theory as well as through two-dimensional, Newtonian, viscous hydrodynamic simulations. In this work, the first of two parts, we limit ourselves to the case of a coplanar disk and binary system. In doing so, we have been able to consider multiple configurations of the only free parameter - the mass ratio of the binary. In both analytical and numerical computations, we have modeled the binary as a pair of point masses with a corresponding Newtonian gravitational potential. We have modeled the disk as a non-self-gravitating, locally isothermal fluid with an alpha-type viscosity. Lastly, in both the analytical and numerical studies, we have limited the domain of interest to the exterior of the binary's orbit by excising the inner low-density region from the numerical domain, and by choosing a harmonic decomposition of the potential in the regime of space outside the semi-major axis of the binary, and the behavior of test-particle orbits under the static background of the combined potential for our analytical computations.  

We have numerically investigated the behavior of the two-dimensional density distribution of the viscous and locally isothermal fluid under the influence of the binary gravitational potential. The computations initialized and evolved four primitive variables: the surface mass density, the surface pressure, and the radial and azimuthal components of the velocity. The knowledge of these variables provides sufficient information for us to calculate other astrophysically relevant characteristics of the disk in post-process.

Since we are focused on the density profiles, the most intuitive and natural way to present our output is to plot the azimuthal- and time-averaged density as a function of radius. Together with certain examples of the full two-dimensional distributions, we can draw broad conclusions about the behavior of matter in the disk. To fully characterize the disk, we also studied other significant characteristics, including the locations of the density maxima and their values, the radii where the viscous and dynamical torque densities are equal, and their value at that point, and the radii where the density in the central cavity drops to 10\% of its instantaneous peak value.

In the analytical computations, we reviewed the theory of approximate orbits in the binary potential, including what the Lindblad resonances are and how they have been used, in combination with linearized fluid mechanics, to explain the formation of a circumbinary gap. We have additionally introduced a measure of the Lyapunov timescale for an analytically tractable set of approximate orbits and, in doing so, we have defined a new measure of the gap size. This new measure reflects the rapidity with which instabilities in the linear regime propagate. We find that the new gap sizes agree better with the trends observed in the full numerical simulations. This supports the conclusion that the circumbinary gap is maintained by the effect of the binary potential on the disk over timescales much shorter than previously suggested. Additionally, we conclude that the viscosity plays a minimal role in the formation of the gap, as the gap sizes are determined solely by the characteristics of the binary. Once again, this suggests a different picture than the case of resonant torque dampening, since that effect would depend on the magnitude of the viscous torque density. 

We add that this conclusion reflects the $\alpha$-disk viscosity profile and the values of $\alpha$ that we have studied. Studies have shown that the morphology of the inner regions of the circumbinary disks are still affected by changing the nature of the viscosity (\cite{DOrazio, 2022MNRAS.513.6158D, 2023arXiv230316204D}). Particularly, \cite{DOrazio} introduces an effective viscosity force to the test particle dynamics when studying the stability of test particle orbits. In such a case, the stability of the Lagrange points L4 and L5 (stationary solutions to \eqref{fullEOM}) indicate a strong dependence of the timescale on the viscous force at mass ratios $q < 0.04$. Extending the orbital stability regime developed here to perturbed orbits around the individual point masses could provide an understanding of the kinematics at radial scales where the point masses are present.

\cite{2022MNRAS.513.6158D} simulate large ranges of viscosities, and demonstrate substantial variability of the gap size as they define it. In addition, \cite{2023arXiv230316204D} study the existence of a decoupled regime as indicated in \cite{2005ApJ...622L..93M}, where the gap size is demonstrated to not track the evolution of the binary at effectively $\alpha\chi^2 > 3\times10^{-4}$. We note that since we fix $\chi=0.1$, this latter range corresponds to $\alpha > 0.03$ in our case, where we have shown that the locations of the Lyapunov-timescale- and torque-balance-based estimates for the gap size begin to overlap. We also expect viscosity to be more dynamically important when it is very large. In addition, we note that the definition for the gap sizes in \cite{2022MNRAS.513.6158D} and \cite{2023arXiv230316204D} is the location where the density is $~20\%$ of $\sigma_0$. This reference density will generally be much larger than ours, since the final density maximum is less than the initial maximum, which in turn is less than $\sigma_0$ (the initial maximum density is only $\sim 0.4 \sigma_0$ in our case), and obviously $10\% < 20\%$; the gaps as defined in \cite{2022MNRAS.513.6158D} and \cite{2023arXiv230316204D} will therefore occur at much larger distances from the binary, where viscosity might become more important. Also, the criterion for the decoupling phase used in \cite{2023arXiv230316204D} is different from that of \cite{2005ApJ...622L..93M}; \cite{2023arXiv230316204D} identify the decoupling phase as when the binary separation and the disk radial velocity are comparable as opposed to the comparison of viscous and gravitational wave radiation timescales.

In our followup paper (Paper II), we will extend the parameter space of configurations by introducing a constant inclination angle between the binary and disk planes. We will apply the same tools used here to characterize the binary's potential and the disk's fluid properties. We will also continue to restrict the domain of interest to the exterior of the binary. 

The primary focus of these two studies is to provide a coherent physical description of the properties of the circumbinary disk and its central gap by combining analytical and numerical approaches. To that effect, the first paper accomplishes the task of providing a self-contained overview of both the analytical and numerical approaches that we consider in this two-part study as well as future applications. It then provides an overview of the inconsistencies in the current analytical understanding of circumbinary disks. Finally, we suggest a new approach, the stability of epicyclic orbits, to reconcile these inconsistencies and to more accurately describe the physics governing the behavior of circumbinary disks.

\section{Acknowledgments}

The authors thank Zoltan Haiman and Alex Dittman for their comments and discussions. SM and STM were supported in part by a National Science Foundation CAREER grant, No. 1945130. The authors acknowledge the computational resources provided by the WVU Research Computing Spruce Knob HPC cluster, which is funded in part by NSF EPS-1003907, and the Thorny Flat HPC cluster, which is funded in part by NSF OAC-1726534. This research was supported in part by the National Science Foundation under Grant No. NSF PHY-1748958.

\bibliography{sample63}{}
\bibliographystyle{aasjournal}

\end{document}